\documentclass[10pt]{iopart}
\usepackage{graphicx,caption2,subfigure}
\makeatletter
\newcommand{\singlespacing}{\let\CS=\@currsize\renewcommand{\baselinestretch}{1.0}\tiny\CS}
\newcommand{\doublespacing}{\let\CS=\@currsize\renewcommand{\baselinestretch}{1.25}\tiny\CS}

\begin{document}
\title[Transverse Momentum Spectra of Pions]{Transverse Momentum Spectra of Pions in Particle and Nuclear Collisions
and Some Ratio-Behaviours: Towards A Combinational Approach}
\author{Bhaskar De\dag\footnote[3]{E-mail:bhaskar$\_$r@isical.ac.in}, S. Bhattacharyya\dag\footnote[4]{(Communicating Author) E-mail: bsubrata@isical.ac.in} and P. Guptaroy\ddag}
\address{\dag\ Physics and Applied Mathematics Unit(PAMU),
Indian Statistical Institute, Kolkata - 700108, India.}
\address{\ddag\ Department of Physics, Raghunathpur College,
Raghunathpur 723133, Purulia (WB), India.}

\begin{abstract}
The nature of transverse momentum dependence of the inclusive
cross-sections for secondary pions produced in high energy
hadronic($PP$), hadronuclear($PA$) and nuclear($AA$) collisions
has here been exhaustively investigated for a varied range of
interactions in a unified way with the help of a master formula.
This formula evolved from a new combination of the basic
Hagedorn's model for particle(pion) production in PP scattering at
ISR range of energies, a phenomenological approach proposed by
Peitzmann for converting the results of $NN(PP)$ reactions to
those for either $PA$ or $AA$ collisions, and a specific form of
parametrization for mass number-dependence of the nuclear cross
sections. This grand combination of models(GCM) is then applied to
analyse the assorted extensive data on various high energy collisions.
The nature of qualitative agreement between measurements and
calculations on both the inclusive cross-sections for production of pions, and some ratios
of them as well, is quite satisfactory. The modest successes that
we achieve here in dealing with the massive data-sets are somewhat
encouraging in view of the diversity of the reactions and the very
wide range of interaction energies.\\
\ \\
\noindent {\bf Keywords} \ : \ Inclusive Cross-sections, Heavy Ion
Collisions.
\end{abstract}
\pacs{13.60.Hb, 25.75.-q, 25.40.Ve.}
\submitto{\JPG}
\maketitle
\newpage
\section{Introduction}
\par
The nature of transverse momentum spectra of the secondaries
produced in various high energy nuclear and particle
collisions\cite{McLerran1,Pirner1,Adler1,Wang1,Wong1,Zhen1} is
still of much importance and interest for some various well-known
reasons. Pions constitute the most abundant variety among the
various secondaries produced in high energy collisions which
include particle-particle, particle-nucleus and nucleus-nucleus
interactions. Within the confine of the present work, the word
`particle' refers mainly to proton, because no other hadron, as
either projectile or target, has here been reckoned with. In this
work we would concentrate on analyzing the nature of transverse
momentum dependence of the inclusive pion production cross section
for some high energy nucleon- and nucleus-induced reactions and
also on understanding the characteristics of some of the cross
section-ratio versus transverse momentum($p_T$) plots. This
indicates certainly no preferential treatment to $p_T$-spectra as
such; rather it is because of the fact that the rapidity spectra
of pions have already been studied with an identical approach, and
the encouraging results of our investigation have been
communicated in a separate work\cite{De1}. The ratio-behaviours
are significant, mainly because of the fact that they offer a
dependable consistency check-up of the final working formula which
has here been framed and forwarded mostly from the
phenomenological points of view; and which has also finally been
put into use for actual calculations of the results. Besides,
comparing the results obtained by the model of our choice with
some of the existing and the most prominent models would also be one of our objectives here. Still, trying to obtain first
reliably good fits to the data on inclusive cross sections versus
transverse momenta ($p_T$) values of the pions produced in $NN$,
$NA$, and $AA$ interactions from a very generalized approach
constitutes here our topmost concern and priority in this work.
\par
Regarding the very basic motivations of this work, we offer
precisely the following few points: (i) The high energy
physicists(both theorists and experimentalists) have, by and
large, focused heavily on $m_T$-scaling behaviour put forward by
Hagedorn for $PP$ collision in his statistical
model\cite{Hagedorn1}. But for production of particles at
relatively high transverse momenta the $m_T$-scaling hypothesis is
seen to be experimentally inapplicable for which Hagedorn himself
later shifted to a power law form\cite{Hagedorn2} of expression
which would be used in the present work. The universal function in
the Hagedorn model with $m_T$-scaling is a Bose-Einstein or Fermi
Dirac distribution and depends on a simple parameter, i.e., the
slope parameter. Contrary to this picture we would try to attack
the problem in reverse and try to determine the universal
behaviour with no particular concern about the status of
$m_T$-scaling by analysing intensively and extensively the
experimental data on both the nucleon-nucleon($PP$) and
nucleon(nucleus)-nucleus interactions. This aspect is unique in
our approach. (ii) Secondly, we have attempted to show somewhat
clearly the degree of dependence of $AA$ collisions on $PP$
collisions and also the extent of intertwining between
proton-proton and nucleus-nucleus reactions. In fact, the entire
energy-dependence of particle production characteristics of
nucleus-nucleus collisions has here been transmuted to the case of
proton-proton reaction at the same center of mass energy. This has
helped us to construct a pathway between $NN$ and $NA/AA$
interactions. (iii) Thirdly, our approach to the analysis of the
problem is not based on any apriori assumptions of the standard
and textual premises of the QGP hypotheses or about the
constituent picture of any of the popular brand of theories.
\par
The organization of this paper is as follows. In the next section
(Section 2) we give the outline of the basic outlook and the
approach to be taken up for this study. In Section 3 we present the essential steps in the methodology of our work here. The Section 4 contains the results of model-based calculations and some broad discussion on the results obtained by this combinational approach. In Section 5 we have made a comparison of the results obtained by the model of our choice and those by some other very front-ranking ones. The last section is reserved for summing up the conclusions and making some points of observational interest and general validity.

\section{The Basic Approach : Tracing the Outline}
\par
Following the suggestion of Faessler\cite{Faessler1} and the work
of Peitzmann\cite{Peitzmann1} and also of Schmidt and
Schukraft\cite{Schmidt1}, we propose here a generalized empirical
relationship between the inclusive cross-section for pion
production in nucleon(N)-nucleon(N) collision and that for
nucleus(A)-nucleus(A) collision as given below:

\begin{equation}
\displaystyle{E\frac{d^3  \sigma}{dp^3} ~ (AB \rightarrow \pi X) ~
\sim ~ (A.B)^{\phi(y, ~ p_T)} ~ E\frac{d^3  \sigma}{dp^3} ~ (PP
\rightarrow \pi X) ~,} \label{e1}
\end{equation}

where $\phi(y, ~ p_T)$ could be expressed in the factorization
form, $\phi(y, ~ p_T) = f(y) ~ g(p_t)$.
\par
While investigating a specific nature of dependence of the two variables
($y$ and $p_T$) either one of these is assumed to remain constant. In
other words, more particularly, if and when the $p_t$-dependence is
studied by experimental groups, the rapidity factor is treated to
be constant and the vice-versa. So, the formula turns into

\begin{equation}
\displaystyle{E\frac{d^3 \sigma}{dp^3} ~ (AB \rightarrow \pi X) ~
\sim ~ (A.B)^{g(p_T)} ~ E\frac{d^3  \sigma}{dp^3} ~ (PP
\rightarrow \pi X) ~,} \label{e2}
\end{equation}

The main bulk of work, thus, converges to the making of an appropriate
choice of form for $g(p_T)$. And the necessary choices are to be made
on the basis of certain premises and physical considerations which do
not violate the canons of high energy particle interactions.
\par
The expression for inclusive cross-section of pions in
proton-proton/antiproton scattering(see figures 1 and 2) at high energies in Eqn.(2) could be chosen in the form suggested first by Hagedorn\cite{Hagedorn2}:
\begin{equation}
\displaystyle{ E\frac{d^3  \sigma}{dp^3} ~ (PP \rightarrow \pi X)
~ = ~ C_1 ~ ( ~ 1 ~ + ~ \frac{p_T}{p_0})^{-n} ~ ,} \label{e3}
\end{equation}

where $C_1$ is the normalization constant, and $p_o$, $n$ are
interaction-dependent chosen phenomenological parameters for which
the values are to be obtained by the method of fitting. Their $\sqrt{s}$-dependences are here proposed to be given by 
the following formulations:

\begin{equation}
\displaystyle{ p_0(\sqrt{s}) ~ = ~ a ~ + ~ \frac{b}{\sqrt{s} ~
\ln({\sqrt{s}})}}
\label{e4}
\end{equation}

where $a ~ = ~ 1.5$, $b ~ = 79.4$ and

\begin{equation}
\displaystyle{ n(\sqrt{s}) ~ =  ~ \acute{a} ~ + ~
\frac{\acute{b}}{\ln^2({\sqrt{s}})}}
\label{e5}
\end{equation}

with $\acute{a} ~ = ~ 6.5$, $\acute{b} ~ = ~ 127$.
 The $\sqrt{s}$-dependence of $p_0$ and $n$ would be shown later diagrammatically in the text and their data-base would also be indicated. The nature and significance of these parameters could be appreciated from the work of Hagedorn\cite{Hagedorn2}, and those of Bielich et al\cite{Bielich1} and Albrecht et al\cite{Albrecht1}. The ratio $p_0/n$ characterizes the intercept of the slope of the transverse momentum spectra in the limit $p_T\rightarrow 0$ and has a value near 150 MeV for $PP$ and also for some nuclear collisions. The prescribed limit of its value is $130-170$ MeV which we have shown to be correct and consistent in our work. 
\par
The final working formula for the nucleus-nucleus collisions is
now being proposed here in the form given below:

\begin{equation}
\displaystyle{E\frac{d^3  \sigma}{dp^3} ~ (AB \rightarrow \pi X) ~
\approx ~ C_2 ~ (A.B)^{(\epsilon ~ + ~ \alpha  p_T  ~ - ~ \beta
p_T^2)} ~ E\frac{d^3 \sigma}{dp^3} ~ (PP \rightarrow \pi X) ~,}
\label{e8}
\end{equation}

with $ g(p_T) ~ = ~ (\epsilon ~ + ~ \alpha  p_T  ~ - ~ \beta
p_T^2)$. This quadratic parametrization is our own suggestion and is called De-Bhattacharyya parametrization(DBP). In the above expression $C_2$ is the normalization term
which has a dependence either on the rapidity or on the rapidity
density of the pion; $\epsilon$, $\alpha$ and $\beta$ are
constants for a specific set of projectile and target.
\par
Earlier works\cite{Albrecht1,Antreasyan1,Aggarwal1} showed that $g(p_T)$ is less than unity in the $p_T$-domain,
$p_T<1.5$ GeV/c. Besides, it was also observed that the parameter
$\epsilon$, which gives the value of $g(p_T)$ at $p_T=0$, is also
less than one and this value differs from collision to collision.
The other two parameters $\alpha$ and $\beta$ essentially
determine the nature of curvature of $g(p_T)$. However, in the
present context precise determination of $\epsilon$ is not
possible for the following understated reasons:
\par
(i) To make our point let us recast the expression for (6) in the form given below:
\begin{equation}
\displaystyle{E\frac{d^3\sigma}{dp^3}(AB \rightarrow \pi X) ~
\approx ~ C_2 ~ (A.B)^\epsilon ~ (A.B)^{(\alpha p_T - \beta
p_T^2)} ~ ( ~ 1 ~ + \frac{p_T}{p_0} ~ )^{-n}}\label{eqn9}
\end{equation}
Quite obviously, we have adopted here the method of fitting.
Now, in Eqn.(7) one finds that there are two constant
terms $C_2$ and $\epsilon$ which are neither the coefficients nor
the exponent terms of any function of the variable, $p_T$. And as
$\epsilon$ is a constant for a specific collision at a specific energy, the product of the two terms $C_2$ and $(A.B)^\epsilon$ appears as just a new constant. And, it is quite difficult to obtain fit-values simultaneously for two constants of the above types through the method of fitting.
\par
(ii) From Eqn.(2) the nature of $g(p_T)$ can easily be determined
by calculating the ratio of the logarithm of the ratios of nuclear-to-$PP$ collision and the logarithm of the product $AB$.
Thus, one can measure $\epsilon$ from the intercept of $g(p_T)$
along y-axis as soon as one gets the values of
$E\frac{d^3\sigma}{dp^3}$ for both $AB$ collision and $PP$
collision at the same c.m. energy. In the present study we have
tried to consider the various collision systems in as many number as
possible. To do so, we have to consider the data on normalized
versions of $E\frac{d^3\sigma}{dp^3}$ for many collision systems
for which clear $E\frac{d^3\sigma}{dp^3}$-data were not available
to us. Furthermore, from these normalized versions we can/could
not extract the appropriate values of $E\frac{d^3\sigma}{dp^3}$ as
the normalization terms, total inclusive
cross-sections$(\sigma_{in})$ etc., for these collision systems cannot always be readily obtained. Besides, it will also not be possible to get readily  the data on inclusive spectra for $PP$ collisions at all c.m.energies, like e.g., at $\sqrt{s}=17.8 GeV$(c.m. energy of $Pb+Pb$ collision).
\par
In order to sidetrack these difficulties and also as an escape-route, we have concentrated almost wholly to the values of $\alpha$ and $\beta$ for various collision systems and the effect
of $C_2$ and $\epsilon$ has been combined into a single constant
term $C_3$. Hence, the final expression becomes,
\begin{equation}
\displaystyle{E\frac{d^3\sigma}{dp^3}(AB \rightarrow \pi X) ~
\approx ~ C_3 ~ (A.B)^{(\alpha p_T - \beta p_T^2)} ~ ( ~ 1 ~ +
\frac{p_T}{p_0} ~ )^{-n}}\label{eqn11}
\end{equation}
with $C_3 = C_2 (A.B)^\epsilon$.
\par
The exponent factor term $\alpha p_T - \beta p_T^2$ obviously represents here $[g(p_T)-\epsilon]$ instead of $g(p_T)$ alone. Thus, after obtaining
fit-values of $\alpha$ and $\beta$, if $[g(p_T)-\epsilon]$ are plotted
for various collision systems, all the curves would originate from
a single point, i.e. origin; and the systems and processes are then really comparable. In other words, in this convenient way we could study and check the scaling characteristics of $g(p_T)$ with respect to the collision systems.
\par
The expression(8) given above is the physical embodiment of what we have termed to be the grand combination of models(GCM) that has been
utilized here. The results of $PP$ scattering are
obtained in the above on the basis of eqn.(3) provided by Hagedorn's
model(HM);  and the route for converting the results of $NN$ to
$NA$ or $AB$ collisions is built up by the Peitzmann's approach(PA) represented by expression(2). The further input is the De-Bhattacharyya parametrization for the nature of the exponent. Thus, the GCM
is the totality and resultant of HM, PA and the DBP, all of which are used here.

\section{The Procedure}
At the very start let us mention a few points of special concern.
Though pions are produced in all three varieties in the physical
processes, we have treated their cases here in an average and
charge-independent way. Secondly, we treated theoretically the
$PP$ and $P\bar{P}$ reactions at accelerator or collider energies
on an equivalent footing, as annihilation channels are found to
have no or negligible contributions at high energies,
$\sqrt{s}\geq10$ GeV. Besides, while collecting data-sets we have,
at times, assumed and used the total negative particle
multiplicity to be physically equivalent to negative pion
multiplicity for all practical purposes, as the pions are
overwhelmingly predominant in number compared to all other
particle-species. And this is done only under compulsion arising
out of the non-availability of direct and reliable data on pion
production in the relevant collision.
\par
The first step towards progress in the present work consists of
fitting the inclusive transverse momentum cross sections of pions
produced in both $PP$ and $P\bar{P}$ collisions available at
various high energies with the help of the proposed form given in
Eqn.(3). The values of $p_0$ and $n$ at different high energies
are specifically noted in Table 1 and Table 2 for the case of
pion production alone. Next, we draw graphs to investigate closely
the nature of $\sqrt{s}$-dependence of these two model-based
parameters, viz, $p_0$ and n. The plots of $p_0$ vs. $\sqrt{s}$
and $n$ vs. $\sqrt{s}$ are presented in Fig.3. While plotting
graphs we have obviously assumed that at very high energies the
proton-proton and proton-antiproton collisions could be treated at
par and with the acceptance of equivalence between each other.
\par
Hereafter, in studying the nature of $p_T$-spectra in all
nucleon-nucleus and nucleus-nucleus collisions, the interaction
energy in all cases of nucleus-induced reactions is invariably
converted first into the c.m. system values, that is expressed in
$\sqrt{s_{\rm NN}}$, then the values of $p_0$ and $n$ are picked
up from the already drawn graphical plot shown by Fig.3. While
analyzing nucleus-dependence with expression(8) and trying with
the fit parameters $C_3$, $\alpha$ and $\beta$, we have inserted
these extracted values of $p_0$ and $n$ for $PP$ cross section
term occurring in the expression(8) given above. The values of
$C_3, ~ \alpha, ~ \beta$, $\chi^2/$ndf and some other relevant information on the specifities for various
nucleon-nucleus and nucleus-nucleus collisions are depicted in
tabular form [Table-3]. The solid curves in all the diagrams
represent the theoretical calculations. The fits, in most cases,
are done within the $p_T$-band $0.8 \leq p_T \leq 3.0$ GeV/c. And this
choice is obviously attuned to the availability of the experimental
data in the several nuclear collisions. However, only exceptions are the cases of $DD$ and $\alpha \alpha$ interactions at $E_{Lab}=480A$ GeV where the $p_T$-ranges are much wider. And this would be evident from the Fig.5. The ratio values are calculated simply by dividing the valid
expression(8) for the corresponding collisions between any pair of projectile and target with some multiplicative factors shown in the several ratio-figures. No further complications about them are taken here into account.

\section{Model-Based Results and Discussions}
Taking the cue from the preceding section one might state that the actual start of the present work takes place from
plotting of the $p_0$ vs. $\sqrt{s}$ graph, and of $n$ vs.
$\sqrt{s}$ graph(Fig.3), which offer us the necessary lever to
apply the connective corridor or connector built up by Peitzmann's approach(PA) for converting the results of nucleon-nucleon collision to that of nucleon-nucleus or nucleus-nucleus interaction at high energy. The
total work is the combination of these two models and of the
parametrization referred to as DBP. The nature of agreement here
between all model-based calculations and experimental measurements
is modestly fair. The reproductions of the data on invariant
inclusive cross section vs. transverse momentum plots, right from
the proton-deuteron collision to the heaviest lead-lead
interaction, and also the data involving some Uranium-induced
collisions as well, are presented in the diagrams from Fig.4 to
Fig.9. The numerical values shown in the seventh column of Table 3
depict the corresponding $\chi^2/$ndf values which indicate the
quantitative measure of the quite satisfactory nature of fits in
qualitative terms, save perhaps the case of sulphur-lead
reaction shown in Fig.8. Let us treat this as a minor exception for which
various reasons might be at play in the various stages of
measurements. But the diagrams shown in Fig.10 merit special
attention. The parameters $\alpha$ and $\beta$ separately
demonstrate a steadily flat nature at values around $\alpha=0.18\pm 0.03$
and $\beta=0.035\pm 0.009$ respectively with regard to the product of $AB$.
This ensures a saturation for the penetrability of projectile into
particle/parton-structure of the target nuclei. The
energy($\sqrt{s}$)-dependence has already been inserted and there
should be no double counting. Now follow some specific comments on RHIC(BNL)-spectra which claim, in recent times, a special status.  
\par
The latest stir in the field of high energy particle physics is
caused by the spurt of particle production in $AuAu$ collisions at
$\sqrt{s_{NN}}=130$ GeV in the RHIC(BNL). We have concentrated here on the
inclusive cross section of production of neutral pions in the
minimum bias events\cite{David1}, and also for both peripheral and central
collisions\cite{Adcox1}. The values of $\alpha$ given in the fifth column
of Table-3 for $AuAu$ collisions do not show any appreciable
change in magnitude for the three separate cases of the
event-sets, while those for $\beta$ remain same within errors. So,
within the limits of $0.8\leq p_T \leq 3.0$ GeV/c, i.e. within the
clear domain of `soft' particle production the results on $AuAu$
collisions at RHIC experiments modestly corroborate our claim that
the parameters, $\alpha$ and $\beta$ are independent of the
collision centrality. However, with growing values of the
transverse momentum($p_T> 3$ GeV/c) i.e., for hard collision, the
behaviour, in all probability, is likely to turn more complex; and
initial indications to such complications are somewhat manifest by
the slight but detectable deviations of the fits provided by us
here for values around $p_T> 3.5$ GeV/c. So, the present study
does neither essentially contradict the content of Adcox et
al\cite{Adcox1}, nor could essentially support the signal message received from them. Indeed, we have left aside here the measurement on
average productions of charged hadron which obviously include
other varieties of charged mesons and baryon-antibaryons as well.
As their incorporations might spoil the exclusivity of pion
production case, we have purposefully sidetracked those data in
the present work; and we have converged more particularly to the
transverse momentum spectra of only the neutral pions.
\par
The features of the cross section ratios(CSR) have been displayed
in the diagrams from Fig.11 to Fig.16. The plots presented in
Fig.16 on prediction alone with no data are the
exceptional ones. The patterns of the plots of the specific ratios versus the transverse momentum ranges are, in most cases, modestly similar
with exceptions for the cases of $\frac{\alpha +
\alpha}{P+P}$(Fig.12),$\frac{\alpha + \alpha}{D+D}$(Fig.12) and
$\frac{S+U}{O+U}$(Fig.14) wherein the curvatures of both the data
plots and the theoretical curves are just the opposite. But, in so
far as the question of agreement between the theoretical plot
versus data is concerned, there is no problem. The CSR behaviours
in the two particular cases of $\frac{\alpha+\alpha}{D+D}$(Fig.12), $\frac{Au+Au}{P+P}$(Fig.16a) and $\frac{Pb+Pb}{D+D}$(Fig.16b) are also of obverse type, though for the former the nature of fit between theory and experiment is quite satisfactory; and for the latter two this point does not just arise as they are predictive theoretical plots. But, the question of similarity or dissimilarity between any two or among the graphs cannot be delved into in any further detail, as the data-points
shown in the graphs indicate the measurements at very different
energies. The actual data on predicted ratio behaviour in $\frac{Au+Au}{P+P}$ could be obtained separately from the RHIC in future. The nature of the ratio of $\frac{Pb+Pb}{D+D}$ could be of interest in the sense that the deuterons are the lightest known nuclei used in colliding systems and $Pb$ is the available heaviest nuclear species which is nowadays being frequently used as both projectile and target in the CERN SPS and will be used in future in the LHC. The dotted curves in both Fig.16(a) and Fig.16(b) indicate the range of theoretical uncertainties due to errors of averaged $\alpha$ and $\beta$. 
\par
The fair agreement between the data and calculations for the
ratios offers us a modestly reliable and somewhat convincing
cross-check of the model for pion production in nuclear collisions
that has here been proposed and developed to a certain extent.
\par
Now, we hold the view that our analysis and results could be made far more rigorous, should the data on pion production in both $PP$ and $NA/AA$ interactions for different $p_T$ in exactly the same rapidity region would be simultaneously available. But by all indications, such expectations at this stage are too unrealistic for which we have had to remain satisfied with whatever is at hand now. It is very encouraging to note that the fits reproduce data on pion production even in the not-very compatible region of rapidity in $PP$ and $AA$ collisions. So, it is only natural and reasonable to summarize that the same or compatible region of rapidity for both types of collisions could offer only better and much-improved fits.

\section{The $p_T$-Spectra in Nuclear Collision: A Brief Comparison of Some Models}
Based on the analysis of the experimental data, let us mention the following few points as the summary of the main and basic features of the $p_T$-spectra, especially of large-$p_T$ variety, of the secondaries produced in nuclear collisions: (i) strong dependence on both target and projectile mass or mass number, (ii) very weak or no dependence on the rapidity, (iii) very weak or no dependence on the c.m.energy, (iv) dependence on impact parameter for those restricted collision wherein the full overlap of target and projectile does not occur, (v) no further $E_T$ dependence beyond the above point. With considerations on some of these characteristics, the NA34 collaboration\cite{Akesson1} demonstrated that the target and projectile dependence could be treated in a similar manner when parametrized by a power law of the form:
\begin{equation}
\displaystyle{\frac{d\sigma}{dp_T}(P+A) ~ = ~ A^{\xi(p_T)} ~ \frac{d\sigma}{dp_T}(P+P)}
\end{equation}
\begin{equation}
\displaystyle{\frac{d\sigma}{dp_T}(B+A) ~ = ~ B^{\zeta(p_T)} ~ \frac{d\sigma}{dp_T}(P+A)}
\end{equation}
Schmidt and Schukraft\cite{Schmidt1} pointed out how a fit of $\xi(p_T)$ to the original Fermilab $P+A$ data could lead to the reasonably nice description of $\zeta(p_T)$ in a two-step manner. The similarity between $\xi(p_T)$ and $\zeta(p_T)$ was emphasized and was interpreted as dependence of the Cronin effect on the projectile mass number analogous to the one on the target.
\par
Next, the WA80 Collaboration\cite{Albrecht1} made attempt to unify functionally this two-stage transitions into a single-step affair in the form of the eqn.(2) dealt with previously in Section 2. Therein
the expression for $g(p_T)$ will have to be chosen phenomenologically in such an appropriate manner as to fit the data as nicely as possible. The WA80 team proposed a simple linear form of fit for $g(p_T)$. in the present work we have taken a polynomial(quadratic) form instead. There is one more major difference between these two approaches. The WA80 team made choices of $p_0$ and $n$ values for $PP$ and $AB$ reactions completely separately. On the contrary we have introduced the results of $p_0$ and $n$ directly from our studies $PP/P\bar{P}$ reactions at various energies. In the approach presented and pursued here the point is to find out the energy of interactions between the nuclei; and once it is known, the $p_0$ and $n$ values of the $PP$ reactions at this specific energy are to be ascertained from Fig.3 and inducted in the calculations.
\par
The performance by WA80 Collaboration\cite{Albrecht1} is essentially grounded in a sort of the physics of purturbative quantum chromodynamics(PQCD) approach in Hagedorn's form, and is also embedded in string and hydrodynamic considerations. This combination is called here the Mixed Model(MM). So, we represent the calculated results of WA80 collaboration as the product of a Mixed Model(MM). The comparison of GCM and MM with experimental data on $SS$ and $SAu$ reactions is given in Fig.17a. The solid lines in all the figures describe results obtained by GCM. Over the entire $p_T$-range the agreement with this MM is quite satisfactory; and at very high $p_T$-values the results based on MM have shown much better agreement than those obtained with the GCM. But there is a change in the scenario in cases of data interpretation on $PbPb$ and $PbNb$ cases(Fig.17a). The comparison, in this two cases, is with calculations made by Thermal Model\cite{Aggarwal1} which does not show quite fair agreement with data for the region beyond $p_T\approx 2.5$ GeV/c, whereas the GCM addresses competently all the measured data-points. In Fig.17(b) a comparison has been made between two sets of models on data from $AuAu$ collisions with two different detectors, for $PbPb$ interaction and also for $OAu$ reaction. And the sets of the models are (i) Modified Hydrodynamic model of Peitzmann\cite{Peitzmann2} and the GCM and (ii) the PQCD-based HIJING model\cite{Wang2} vs. the GCM. The agreement with both data and the most popular versions of the two models could, on the whole, be rated to be modestly fair. The HIJING model is old and well-known among the Monte-Carlo simulation-based models. And, unlike the case of hydrodynamic model, there has not been any new updating of it in the recent past.
\par
Compared to the GCM, the version of hydrodynamic model offered very recently by Peitzmann\cite{Peitzmann2} provides certainly a little better agreement, especially at large values of $p_T$. This is observed in Fig.17b. The data measured from $AuAu$ reactions with two different detectors and other accessories have been shown separately and the other set of data is from $PbPb$ collision. But the hydrodynamic model of Peitzmann is based on a version of this model of Weidemann and Heinz who includes transverse flow and resonance decays. Besides, Peitzmann also introduced the spatial distribution of Woods-Saxon type instead of simple Gaussian; reckoned with the difference between chemical freeze-out and kinetic freeze-out temperatures instead of one universal freeze-out temperature, and also the role of baryonic chemical potential.  The certain degree of rigour in calculations is reflected in fair achievements in the collisions of the two heavies. But even with such physics considerations for actual calculations, the number of hand-inserted parameters are too large in the Peitzmann's latest model compared to just three in total in our approach. In our case the entire collective effects arising out of multiple nucleon-nucleon collisions and the effects of nuclear geometry are absorbed by the very simple term, $(AB)^{g(p_T)}$. And the rest depends on just the results of $PP$ collisions. So, our approach is much straightforward compared to the existing other numerous complicated approaches of the models.

\section{Summary and Outlook}
What we have done here seems, by now, to be pretty clear. We have precisely
introduced a simple parametrization of pion transverse momentum
spectra for nucleus-nucleus collisions at high energies in terms
of that for proton-proton collision at the same energy. Analyzing
a large variety of colliding systems at different energies we have
shown that the ratio of the two transverse momentum spectra
approximately proportional to the product of the mass numbers of
the colliding systems raised to a power that is a quadratic
function of the transverse momentum of the particle under
consideration. Furthermore, this quadratic function is essentially
universal, i.e. its coefficients have only very weak dependence on the
colliding systems and energies.
\par
Through the massive demonstration of the fair agreements between
measured data and our calculations the merit of the totality of the
present approach is modestly well-made from the functional point
of view. Despite this, let us sum up first here the important
finer physical points which are the distinctive traits of the GCM
alone. (i) Firstly, the GCM provides the most economical
description of the vast body of experimental data on pion
production in diverse set of reactions with only three arbitrary
parameters. (ii) Secondly, the choice of the quadratic
parametrizations for $p_T$-dependence of nuclear cross-sections is
one of our important contributions. (iii) Thirdly, we have suggested
independently an approach for computing the values of inclusive
cross sections for production of pions in $PP$ collision with the
help of the plots on $p_0$ vs. $\sqrt{s}$ and $n$ vs. $\sqrt{s}$ at any high energy point of the c.m.energy values. (iv) Once the energy-dependence in $PP$ collision is taken care of by this method, the
inclusive cross sections for pion production in $NA/AB$
interactions becomes essentially energy-independent. In other
words, energy-dependence in nuclear collisions at high energy is
manifested only through the very basic interacting sets of
proton-proton collisions. No reflection of energy-dependence is
detected in this mechanism in the nuclear geometry of collisions.
This is in striking contrast with the most of the existing models.
(v) We have also arrived at a sort of mass-number scaling of the
parameters in the exponent of the nuclear dependent terms. This is
also in sharp variance with the standard frameworks, though
admittedly such a scenario is possible with only suitable
adjustment of the normalization terms in the inclusive cross sections.
(vi) Besides, we have tried to provide substantial support to
the proposed connector here between $NN$ and $NA/AB$ reactions by making our studies as extensive as possible in order to impart a good degree of reliability to this combinational approach. (vii) Finally, this work 
brings out the features of `universality' of all high energy particle and nuclear interactions in a satisfactory manner.
\par
But the approach, despite its successes, suffers from the following
gross limitations: (i) The master formula that is at the root of the success is not and cannot be derived right now from any of the first principles of basic
physics. (ii) Data on production of a specific variety of
particles in $PP$ collisions at a few high energies are an
indispensable necessity for analyzing and explaining the case of production of this particular particle species in any nuclear collision. This $PP$-dependency is, by all indications, an inseparable part of this grand combination of models (GCM), though this might appear, at times, to be an intractable problem arising out of the lack of measured data on $PP$ reaction at any particular energy. (iii) Thirdly and finally, the approach is totally insensitive to the physics of space-time-evolution of the collisions. But, seen from the angle of calculations on production of the final particles, this might be considered an advantage as well. 


\newpage
i) PP Collision:
\begin{table}[h]
\caption{Fit Values of $p_0$ and $n$ for $PP$ Collisions at
different energies.}
\footnotesize{
\begin{tabular}{@{}lllllll}
\br 
$\sqrt{s}_{NN}(GeV)$ & Relevant collision-specifics & $C_1$ & $p_{0}$ & $n$ & $\frac{\chi^2}{ndf}$ & $p_0/n$ \\
\mr 
$23$ & & $192 \pm 6$ & $2.48 \pm 0.12$ & $18.88 \pm 0.76$ & $1.539$ & 0.131  \\
 $31$ & $\pi^-$, & $219 \pm 5$ & $2.229\pm 0.004$ & $17.08\pm 0.04$ & $1.157$ & 0.130 \\
 $45$ & $y_{cm}=0$, & $241 \pm 5$ & $1.89\pm 0.01$ & $14.97\pm 0.06$ & $1.352$ & 0.126 \\
 $ 53$ & min. bias & $240 \pm 7$ & $1.83\pm 0.01$ & $ 14.25\pm 0.03$ & $2.72$ & 0.128 \\
 $63$ & & $248 \pm 6$ & $1.78 \pm .01 $ & $ 14.07\pm 0.02$ & $1.621$ & 0.127 \\
\br
\end{tabular}}
\end{table}
\bigskip

ii) P$\bar{\rm P}$ Collision:

\begin{table}[h]
\caption{Fit Values of $p_0$ and $n$ for $P \bar{P}$ Collisions at
different energies.}
{\footnotesize{
\begin{tabular}{@{}lllllll}
\br $\sqrt{s}_{NN}(GeV)$ & Relevant collision-specifics& $C_1$
& $p_0$ & $n$ &
$\frac{\chi^2}{ndf}$ & $p_0/n$ \\
\mr $200$ & charged, $\eta_{cm}=0$, min. bias & $414 \pm 7$ &
$1.52\pm 0.01$ & $11.18\pm 0.02$ &
$1.614$ & 0.136\\
$500$ & charged, $\eta_{cm}=0$, min. bias & $327 \pm 7$ & $1.51\pm0.05$ & $9.84\pm 0.03$ & $1.787$ & 0.153\\
$630$ & $(h^++h^-)$, $-3.0<\eta_{cm}<3.0$, min. bias & $302 \pm 5$ & $1.49\pm 0.01$ & $9.41\pm 0.01$ & $1.694$ & 0.158\\
$900$ & charged, $\eta_{cm}=0$, min. bias & $412 \pm 9$ & $1.50\pm 0.05$ & $9.43\pm 0.02$ & $1.61$ & 0.159\\
$1800$ & charged, $-1.0<y_{cm}<1.0$, min. bias & $370 \pm 7$ & $1.48\pm 0.01$ & $ 8.79\pm 0.03$ & $1.694$ & 0.168\\
\br
\end{tabular}}}
\end{table}
\newpage
\begin{table}
\caption{Numerical Values of the parameters: $\alpha$ and $\beta$
for different high energy collisions.}
{\footnotesize{
\begin{tabular}{@{}llllllll}
\br  ${\rm Collision}$ & $E$(GeV) & Relevant & $C_3$ &  $\alpha$ & $\beta$ & $\chi^2$/ndf & Fig. \\
 &  & collision-specifics &  &(GeV/c)$^{-1}$  & (GeV/c)$^{-1}$ & & No. \\
\mr ${\rm P+D}$ & $400$ & $\pi^-$, min. bias & $400 \pm 20$ & $0.21 \pm 0.06 $ & $0.04 \pm 0.01$& 0.005 & 4\\
 ${\rm P+Be}$ & $400$ & $\pi^-$, min. bias & $1700 \pm 80$ & $0.20 \pm 0.05 $ & $0.040 \pm 0.015$& 0.008 & 4\\
 ${\rm P+Ti}$ & $400$ & $\pi^-$, min. bias & $(4.5 \pm 0.5)\times10^3$ & $0.21 \pm 0.05$ &  $0.04 \pm 0.015$ &  0.002 & 4\\
 ${\rm P+W}$ & $200$ & $\pi^-$, min. bias & $(3.8 \pm 0.5)\times10^3$ & $0.20 \pm 0.04$ &  $0.04 \pm 0.02$ &  1.624 & 4\\
 ${\rm P+W}$ & $300$ & $\pi^-$, min. bias & $(4.7 \pm 0.3)\times10^3$ & $0.17 \pm 0.02$ &  $0.031 \pm 0.002$ &  1.15 & 4\\
 ${\rm P+W}$ & $400$ & $\pi^-$, min. bias & $(9.4 \pm 0.6)\times10^3$ & $0.14 \pm 0.03$ &  $0.02 \pm 0.01$ &  0.013 & 4\\
 ${\rm P+Au}$ & $200$ & $\pi^0$, $1.5\leq \eta \leq 2.1$, min. bias & $(1.7 \pm 0.2)\times10^4$ & $0.16 \pm 0.01$ &  $0.04 \pm 0.01$ & 0.570 & 4\\
 ${\rm P+U}$ & $200$ & $(\pi^++K^+)$, $3.0<y<3.9$ & $(3.8 \pm 0.1)\times10^4$ & $0.20 \pm 0.02$ &  $0.032 \pm 0.004$ & 0.903 & 4\\
 ${\rm D+D}$ & $480A$ & $\pi^0$, $|y_{cm}|<0.4$ & $360 \pm 16$ & $0.25 \pm 0.05$ &  $0.026 \pm 0.008$ &  1.438 & 5\\
 ${\rm D+D}$ & $1400A$ & $\pi^0$ & $(24 \pm 4)\times 10^4$ & $0.18 \pm 0.03$ &  $0.04 \pm 0.01$ &  1.627 & 5\\
 ${\rm D+D}$ & $1900A$ & $\pi^0$ & $(28 \pm 4)\times 10^4$ & $0.15 \pm 0.02$ &  $0.04 \pm 0.01$ &  1.627 & 5\\
 ${\rm D+Au}$ & $200A$ & $h^-$, $2.0<y<3.0$, central & $138 \pm 20$ & $0.23 \pm 0.02$ &  $0.04 \pm 0.02$ &  1.720 & 8\\
 ${\rm \alpha + \alpha}$ & $480A$  & $\pi^0$, $|y_{cm}|<0.4$ &  $624 \pm 50$ & $0.27 \pm 0.02$ & $0.026 \pm 0.002$ & 0.015 & 5\\
 ${\rm O+C}$ & $200A$ & $\pi^0$, $1.5\leq \eta \leq 2.1$, min. bias & $(7 \pm 1)\times10^3$ & $0.13 \pm 0.02$ &  $0.07 \pm 0.02$ &  1.475 & 6\\
 ${\rm O+W}$ & $200A$ & negative particles, & $(1.3 \pm 0.4)\times10^3$ & $0.18 \pm 0.02$ &  $0.031 \pm 0.005$ & 1.220 & 6\\
 & & $1.0<y<1.9$, min. bias & & & &\\
 ${\rm O+Au}$ & $200A$ & $\pi^0$, $1.5\leq \eta \leq 2.1$, min. bias & $(9 \pm 1)\times10^4$ & $0.161 \pm 0.004$ &  $0.038 \pm 0.003$ & 0.030 & 6\\
 ${\rm O+Au}$ & $200A$ & $h^-$, $2.0<y<3.0$, central & $(1.2 \pm 0.2)\times 10^3$ & $0.15 \pm 0.02$ &  $0.040 \pm 0.005$ &  1.562 & 8\\
 ${\rm O+U}$ & $200A$ & $(\pi^++K^+)$, $3.0<y<3.9$ & $(5.3 \pm 0.2)\times 10^5$ & $0.180 \pm 0.002$ &  $0.036 \pm 0.002$ & 1.359 & 6\\
 ${\rm S+S}$ & $200A$ & $\pi^0$, $2.1<y<2.9$, min. bias & $(6.1 \pm 0.4) \times 10^4 $ & $0.18 \pm 0.01$ &  $0.046 \pm 0.002$ & 1.603 & 7\\
 ${\rm S+Ag}$ & $200A$ & $h^-$, $2.0<y<3.0$, central & $(1.4 \pm 0.1)\times10^3$ & $0.150 \pm 0.004$ &  $0.04 \pm 0.01$ & 1.790 & 8\\
 ${\rm S+W}$ & $200A$ & negative particles, & $(1.13 \pm 0.08)\times10^3$ & $0.17 \pm 0.03$ &  $0.031 \pm 0.005$ & 1.566 & 7\\
 & & $1.0<y<1.9$, min. bias & & & & \\
 ${\rm S+Au}$ &  $200A$ & $\pi^0$, $2.1<y<2.9$, min. bias & $(2.0 \pm 0.2)\times 10^5$ & $0.19 \pm 0.02 $ &  $0.039 \pm 0.005$ & 1.080 & 7\\
 ${\rm S+Pb}$ & $200A$ & negative particles, & $230 \pm 40$ &  $0.15 \pm 0.05$ &  $0.05 \pm 0.02$ & 2.646 & 8\\
 & & $2<y<4$, central & & & & \\
 ${\rm S+U}$ &  $200A$ & $(\pi^++K^+)$, $3.0<y<3.9$ & $(9.0 \pm 0.5)\times 10^5$ & $0.17 \pm 0.02 $ &  $0.03 \pm 0.01$ & 0.433 & 7\\
 ${\rm Au+Au}$ & $8450A$ & $\pi^0$, $|\eta|<0.35$, min. bias & $120 \pm 40$ &  $0.15 \pm 0.02$ &  $0.036 \pm 0.003$ & $0.449$ & 9\\
 ${\rm Au+Au}$ & $8450A$ & $\pi^0$, $|\eta|<0.35$, central &  $500 \pm 30$ & $0.145 \pm 0.008$ &  $0.036 \pm 0.002$ & $1.524$ & 9\\
 ${\rm Au+Au}$ & $8450A$ & $\pi^0$, $|\eta|<0.35$, peripheral &  $27 \pm 7$ & $0.14 \pm 0.02$ &  $0.036 \pm 0.002$ & $1.836$ & 9\\
 ${\rm Pb+Nb}$ &  $160A$ & $\pi^0$, $2.3<y<3.0$, min. bias &  $(2.7 \pm 0.4)\times10^5$ & $0.17 \pm 0.02$ & $0.036 \pm 0.004$ & 1.614 & 7\\
 ${\rm Pb+Au}$ &  $160A$ & charged pions, &  $1230 \pm 5$ & $0.18 \pm 0.03$ & $0.035 \pm 0.002$ & 1.822 & 8\\
 & & $2.1<\eta<2.6$, central & & & &\\
 ${\rm Pb+Pb}$ &  $160A$ & $\pi^0$, $2.3<y<3.0$, min. bias &  $(9 \pm 1)\times10^5$ & $0.16 \pm 0.03$ & $0.03 \pm 0.01$ & 1.070& 7\\
\br
\end{tabular}}}
\end{table}
\newpage
\begin{figure}
\centering
\includegraphics[width=8cm]{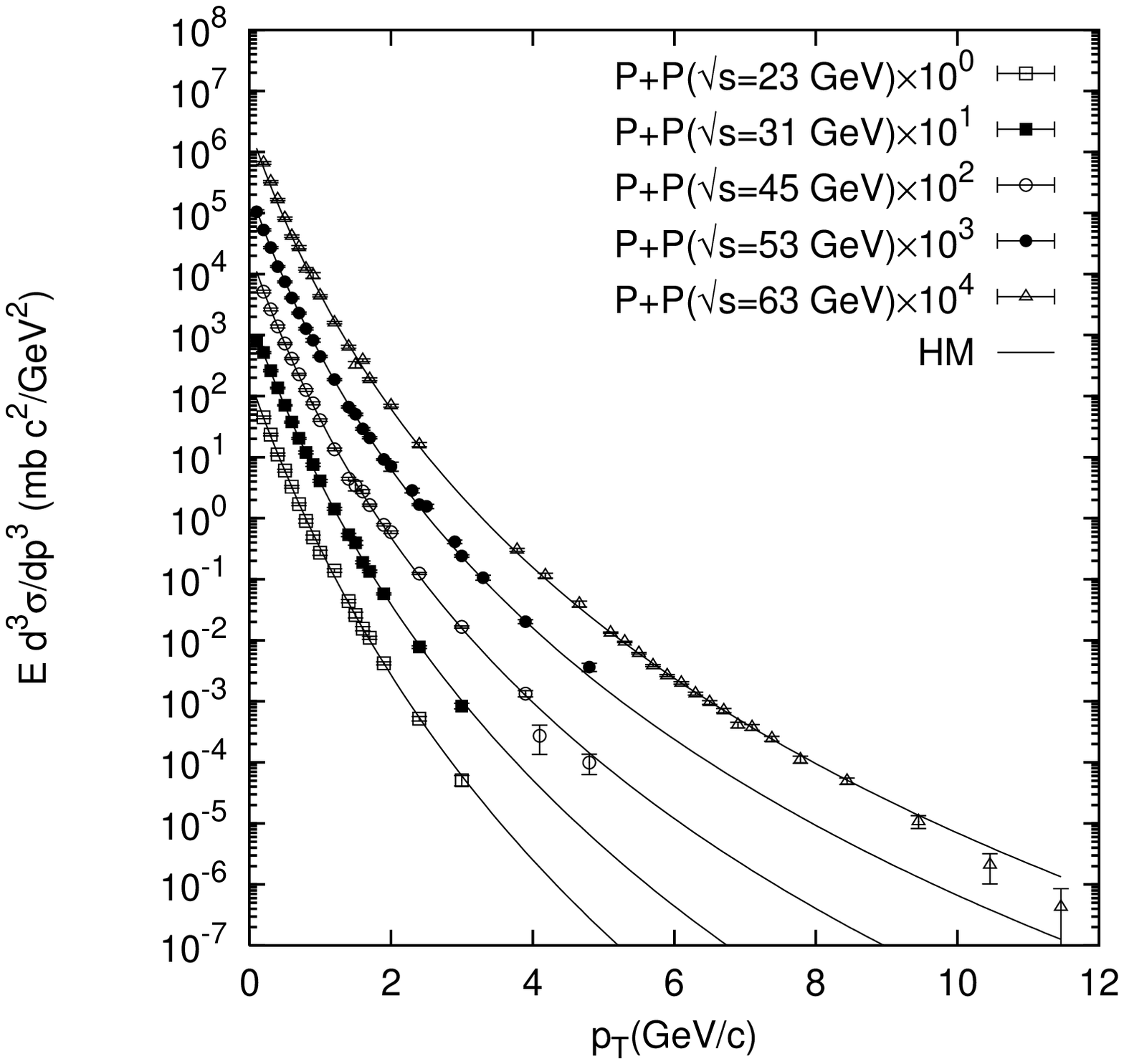}
\label{Fig.1} \caption{Plot of $E\frac{d^3\sigma}{dp^3}$ vs. $p_T$
for secondary pions produced in $P+P$ collisions at different c.m.
energies. The various experimental points are taken from Ref.\cite{Alper1}
and Ref.\cite{Drijard1}. The solid curves give the theoretical fits on the
basis of Hagedorn's model(Eqn.3).}
\includegraphics[width=8cm]{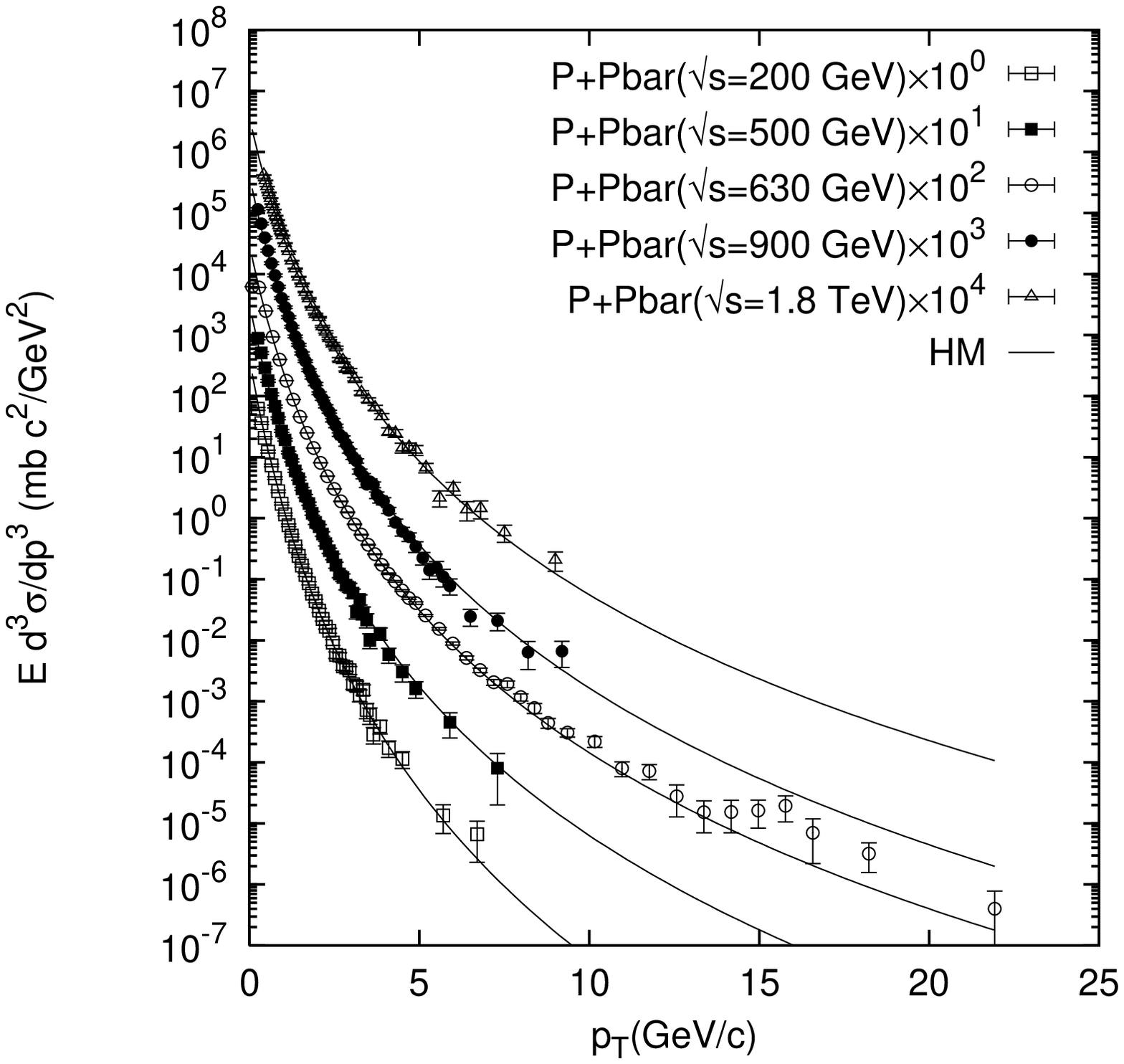}
\label{Fig.2} \caption{ The inclusive spectra for secondary pions
produced in $P+\bar{P}$ collisions at $\sqrt{s}=200,500,630,900$
and $1800$ GeV. The various experimental points are from Ref.\cite{Albajar1},
Ref.\cite{Bocquet1} and Ref.\cite{Abe1}. The solid curvilinear lines are drawn on the
basis of Eqn.3.}
\end{figure}
\begin{figure}
\centering
\includegraphics[width=8.5cm]{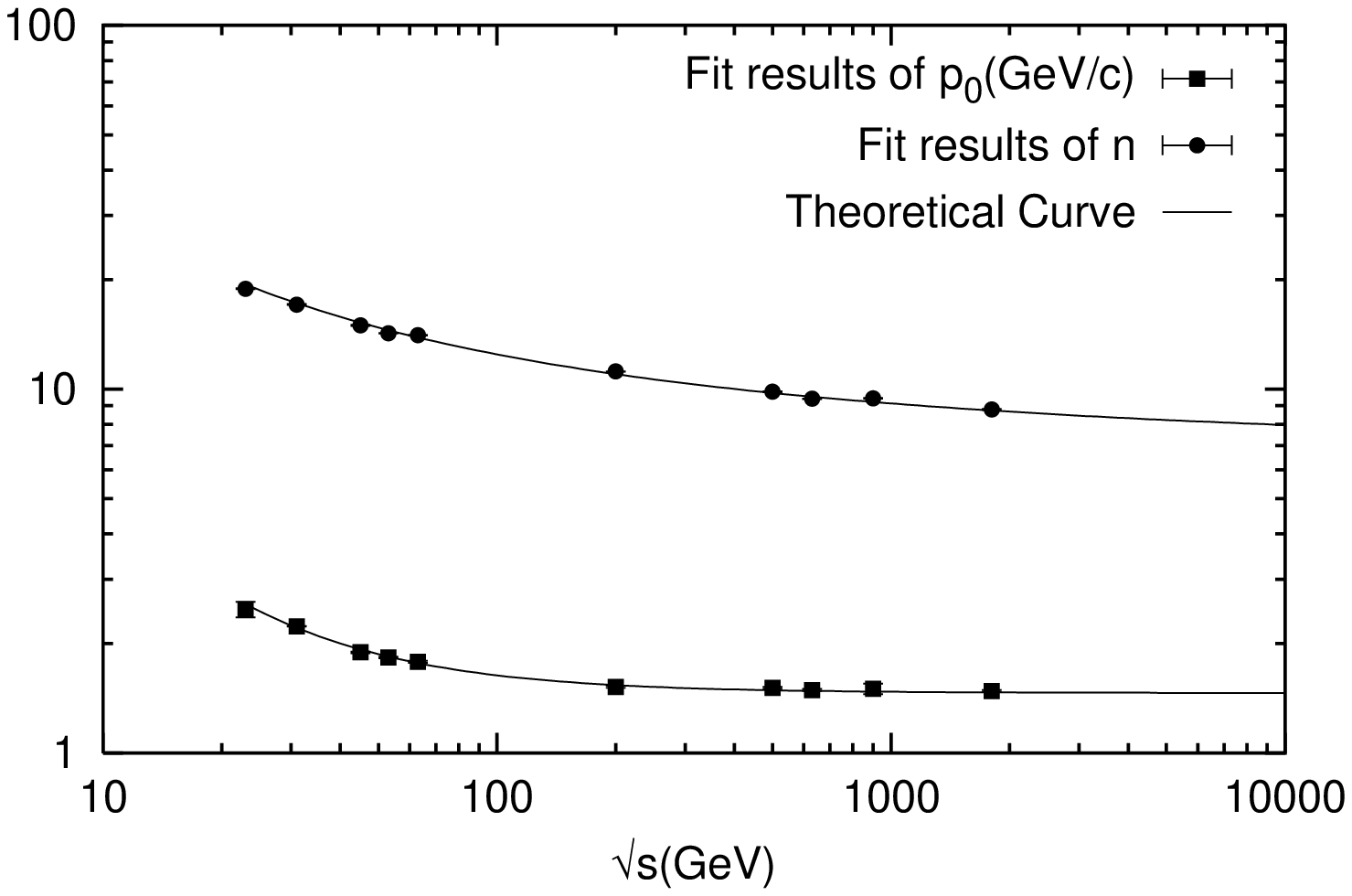}
\label{Fig.3} \caption{Values of $p_0$ and $n$ as a function of
c.m. energy $\sqrt{s}$. Various data points are taken from
Table-1 and Table-2. The solid curves are drawn on the basis
of Eqn.4 and Eqn.5.}
\includegraphics[width=8.5cm]{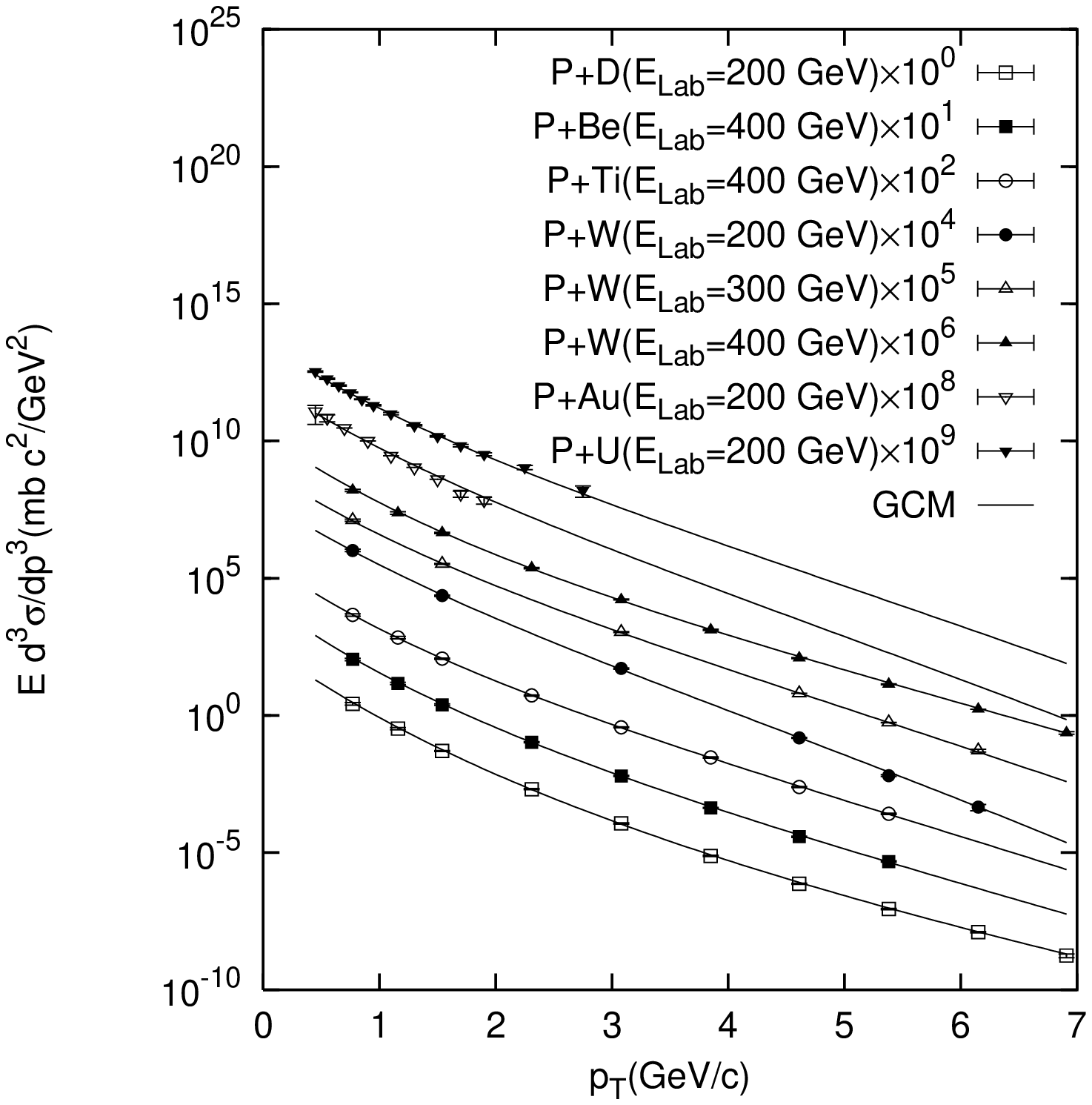}
\label{Fig.4} \caption{Plot of $E\frac{d^3\sigma}{dp^3}$ for
production of secondary pions in some Proton induced collisions at
$E_{Lab}=200,300$ and $400$ GeV as a function of transverse
momentum $p_T$. The various experimental data-points are taken
from Ref.\cite{Antreasyan1}, Ref.\cite{Albrecht2} and Ref.\cite{Abreu1}. The solid curves represent the
fits on the basis of Eqn.8.}
\end{figure}
\begin{figure}
\centering
\includegraphics[width=8.5cm]{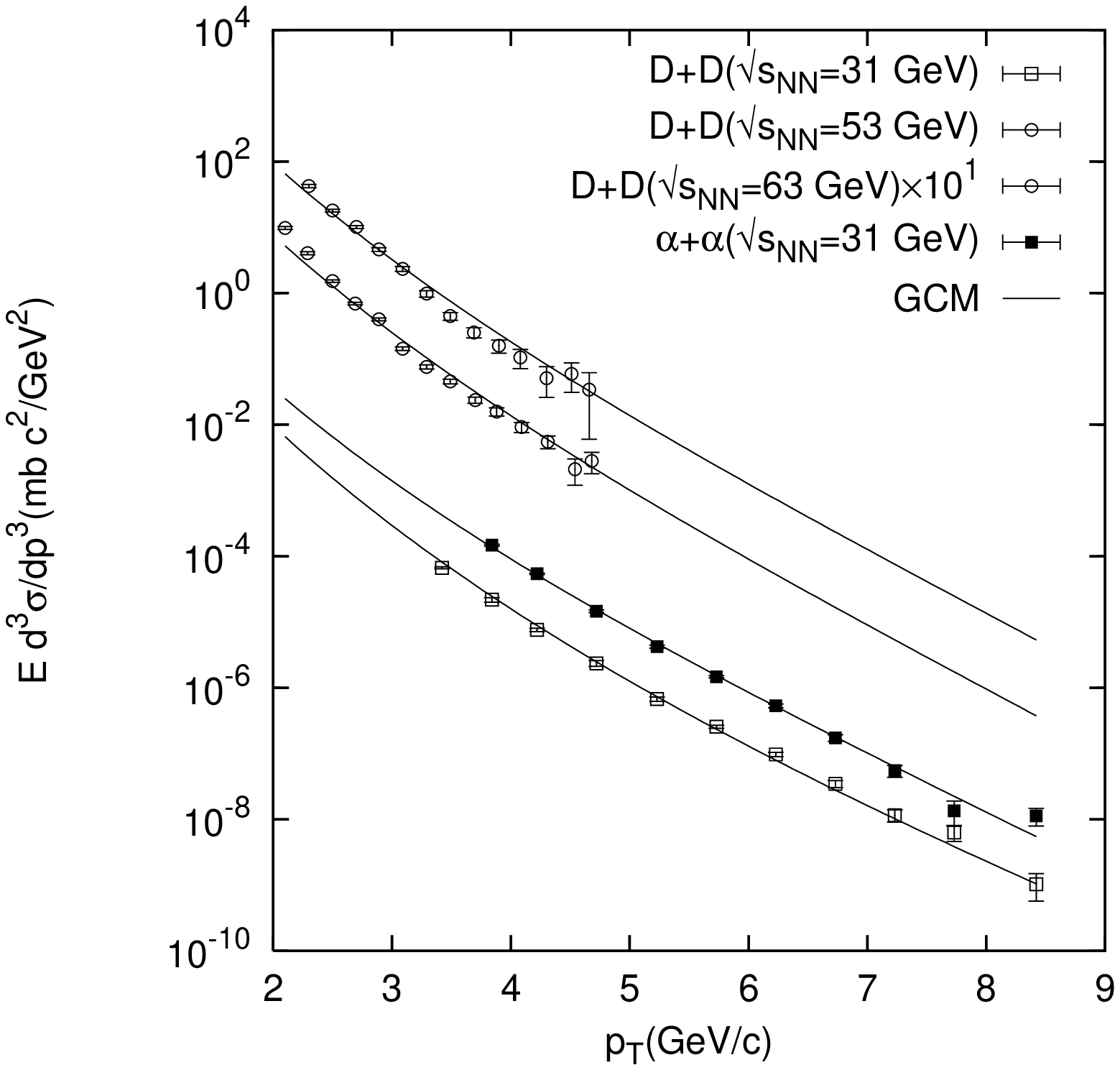}
\label{Fig.5}
 \caption{Plot of
$E\frac{d^3\sigma}{dp^3}$ vs. $p_T$ for secondary pion produced in
$D+D$ and $\alpha+\alpha$ collisions at different energies. The
various experimental points are taken from Ref.\cite{Angelis1} and Ref.\cite{Clark1}. The
solid curves provide the fits on the basis of Eqn.8.}
\includegraphics[width=8.5cm]{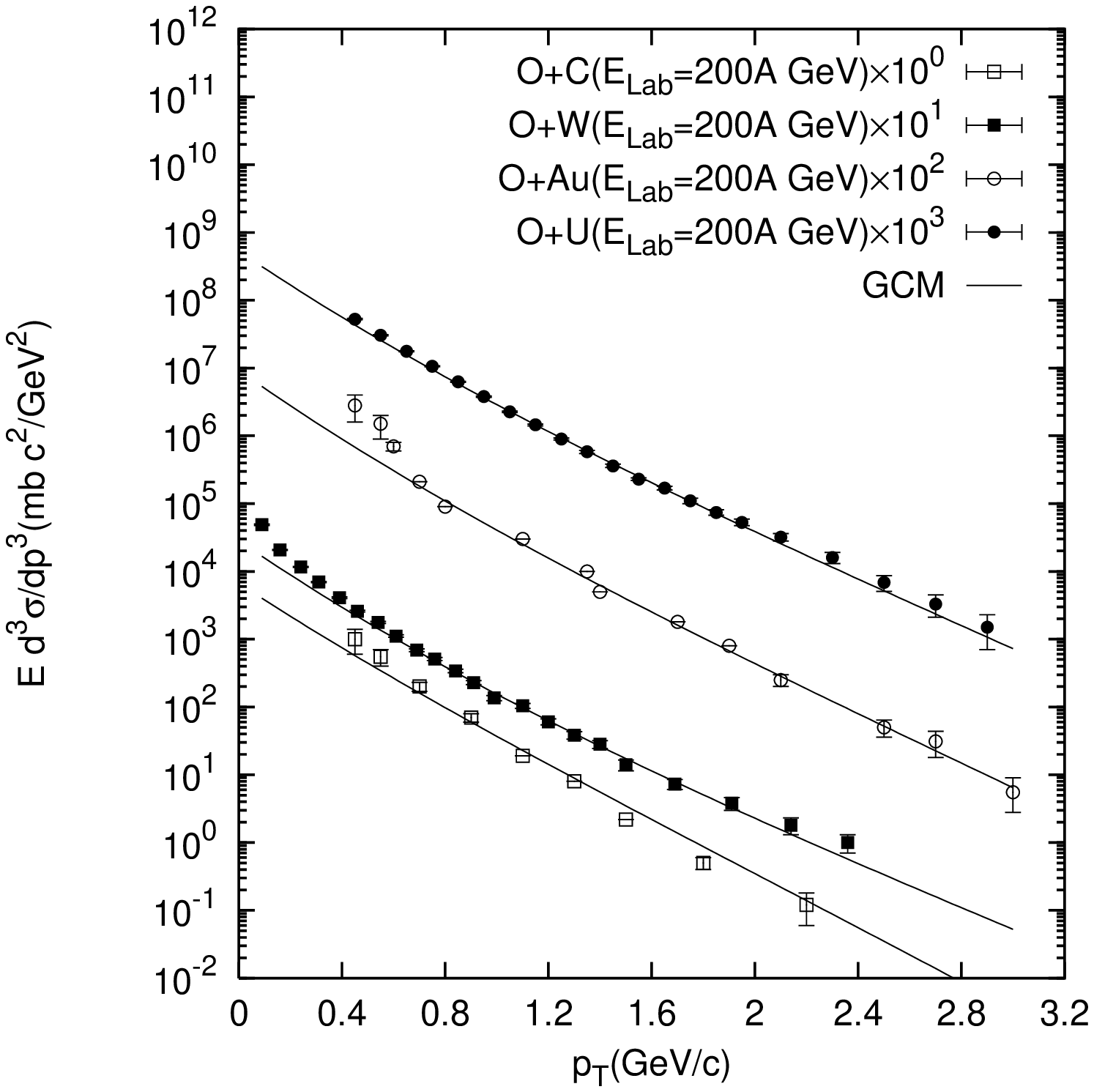}
\label{Fig.6} \caption{Nature of inclusive spectra for production
of secondary pions in different Oxygen induced collisions at
$E_{Lab}=200$A GeV. The various experimental points are taken from
 Ref.\cite{Albrecht2}, Ref.\cite{Abreu1} and Ref.\cite{Akesson2}. The solid curvilinear lines represent
the fits on the basis of Eqn.8.}
\end{figure}
\begin{figure}
\centering
\includegraphics[width=8.5cm]{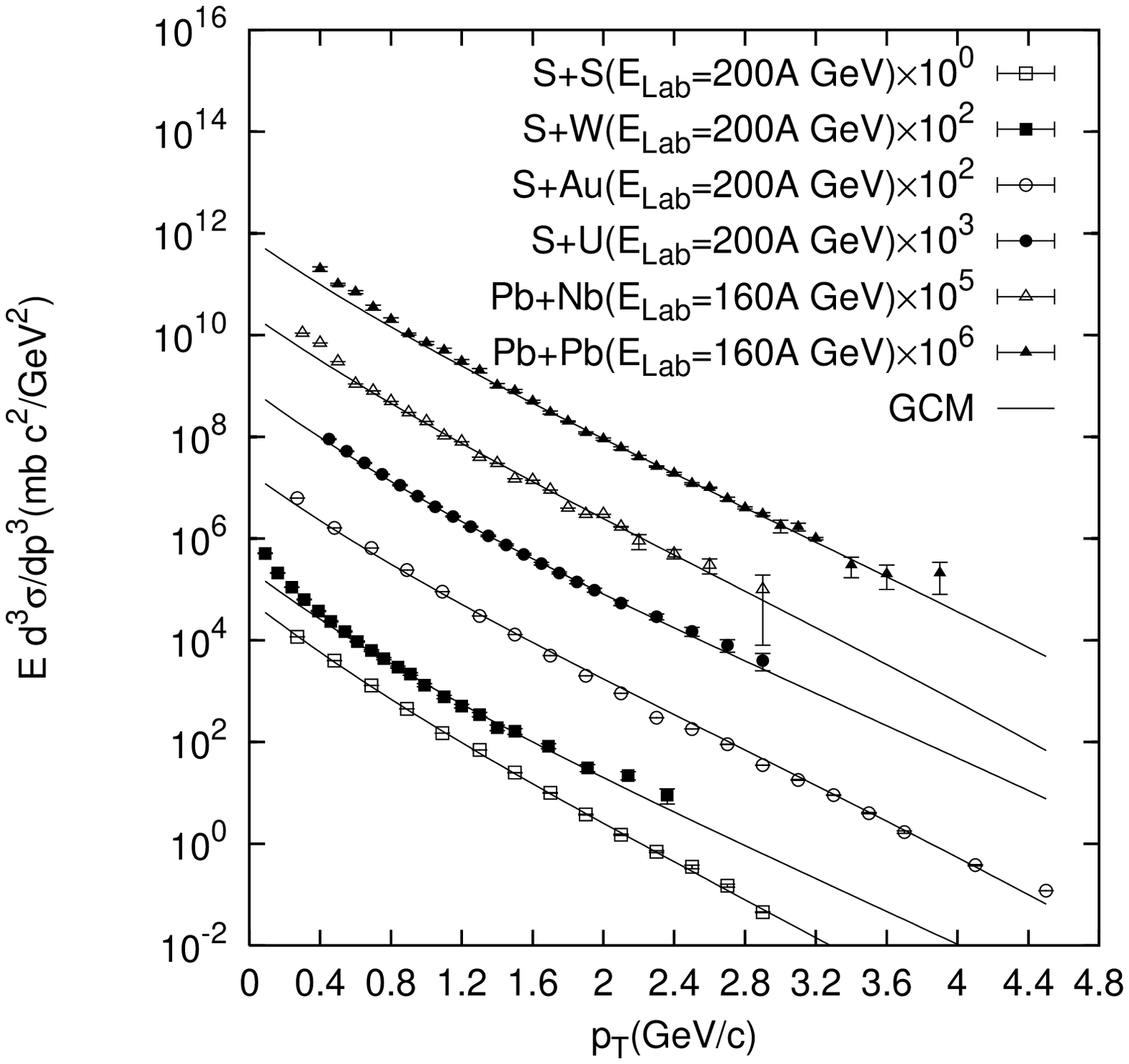}
\label{Fig.7} \caption{Plot of inclusive spectra vs. $p_T$ for
production of secondary pions in some sulphur-induced and lead-induced reactions at $E_{Lab}=200$A GeV and at $E_{Lab}=160$A. The various experimental data-points are from
Ref.\cite{Albrecht1}, Ref.\cite{Aggarwal1}, Ref.\cite{Abreu1} and Ref.\cite{Akesson2}. The solid curves provide the
fits on the basis of Eqn.8.}
\includegraphics[width=8.5cm]{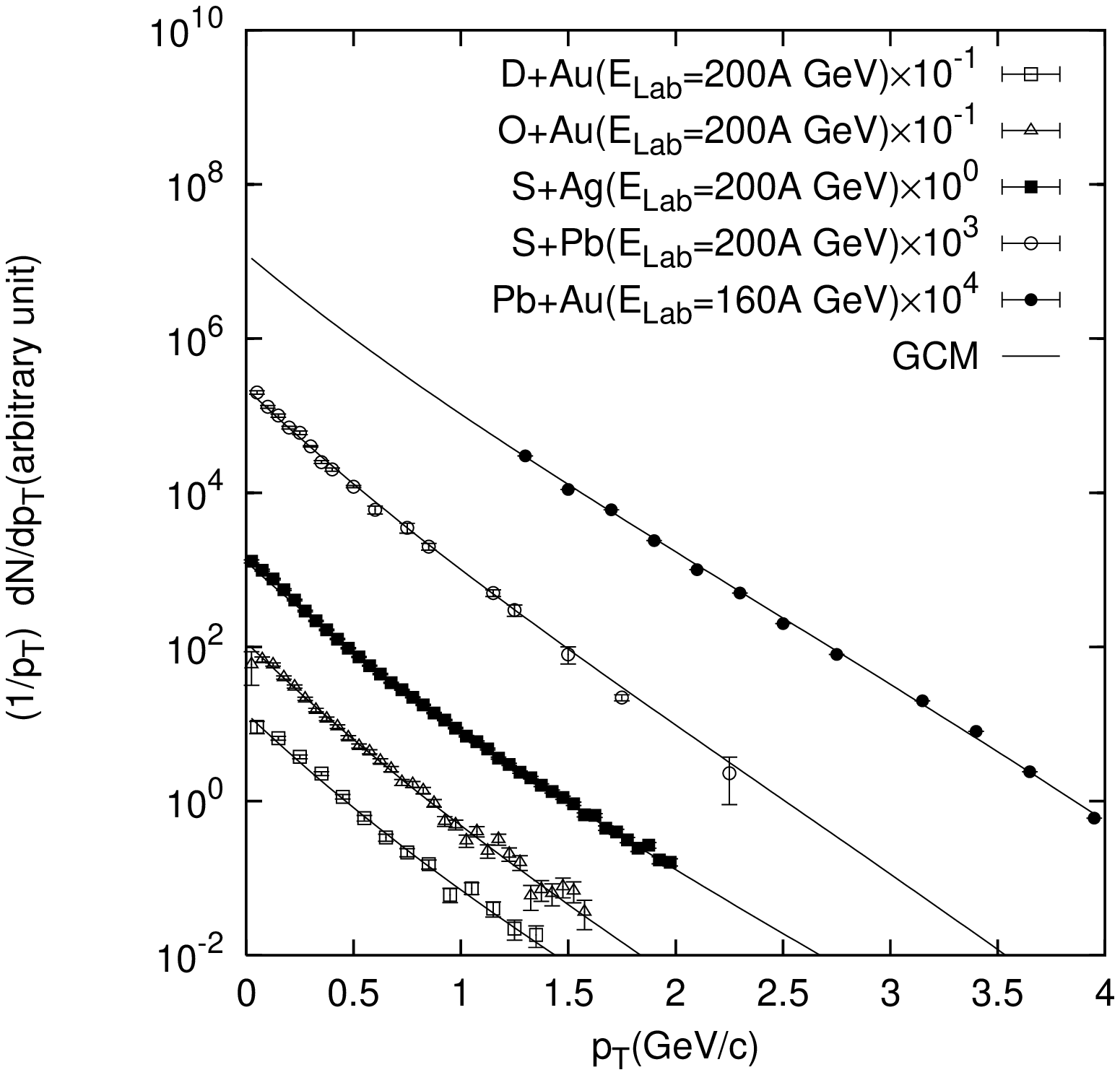}
\label{Fig.8}
 \caption{The nature of $\frac{dN}{dp_T^2}$ for production of secondary
pions in $D+Au$, $O+Au$, $S+Ag$ and $S+Pb$ collisions at
$E_{Lab}=200A$ GeV and in $Pb+Au$ collision at $E_{Lab}=160A$ GeV with change of transverse momentum. The various
experimental points are taken from Ref.\cite{Alber1},
Ref.\cite{Wlodarczyk1} and \cite{Agakichiev1}. The solid curves
represent the fits on the basis of Eqn.8.}
\end{figure}
\begin{figure}
\centering
\includegraphics[width=8.5cm]{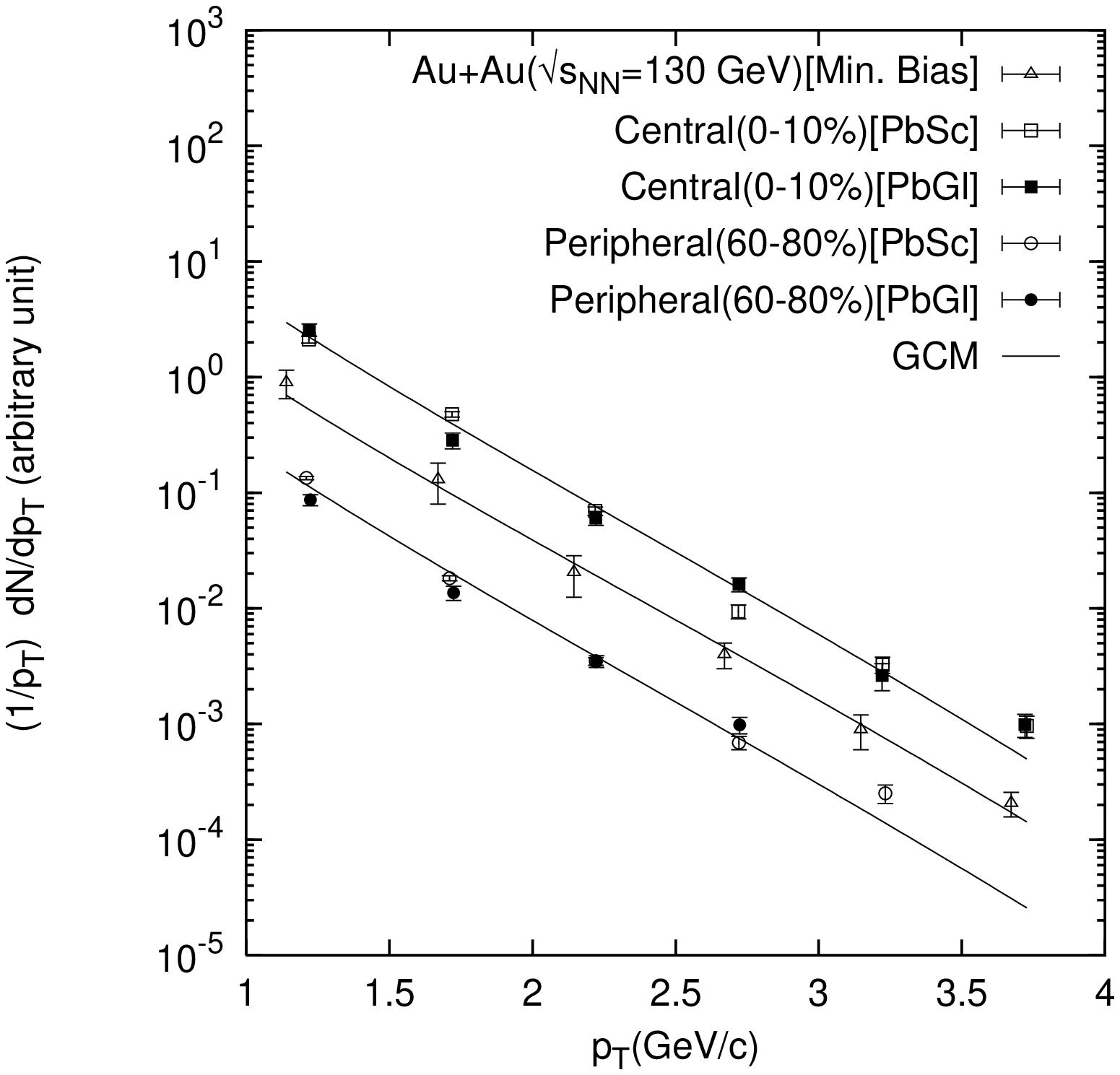}
\label{Fig.9} \caption{Invariant spectra for secondary pions
produced in $Au+Au$ reaction at $\sqrt{s_{\rm NN}}=130$ GeV(RHIC).
The various experimental points are from Ref.\cite{David1} and Ref.\cite{Adcox1}.Against the general background of pion production in $Au+Au$ reaction, the symbols PbSc and PbGl indicate simply the two
separate calorimeters; the former is the presentations of data-set
obtained by lead-scintillator sampling calorimeter and the latter
are those by lead-glass Cerenkov calorimeter. The solid curves
give the fits on the basis of Eqn.8.}
\includegraphics[width=8.5cm]{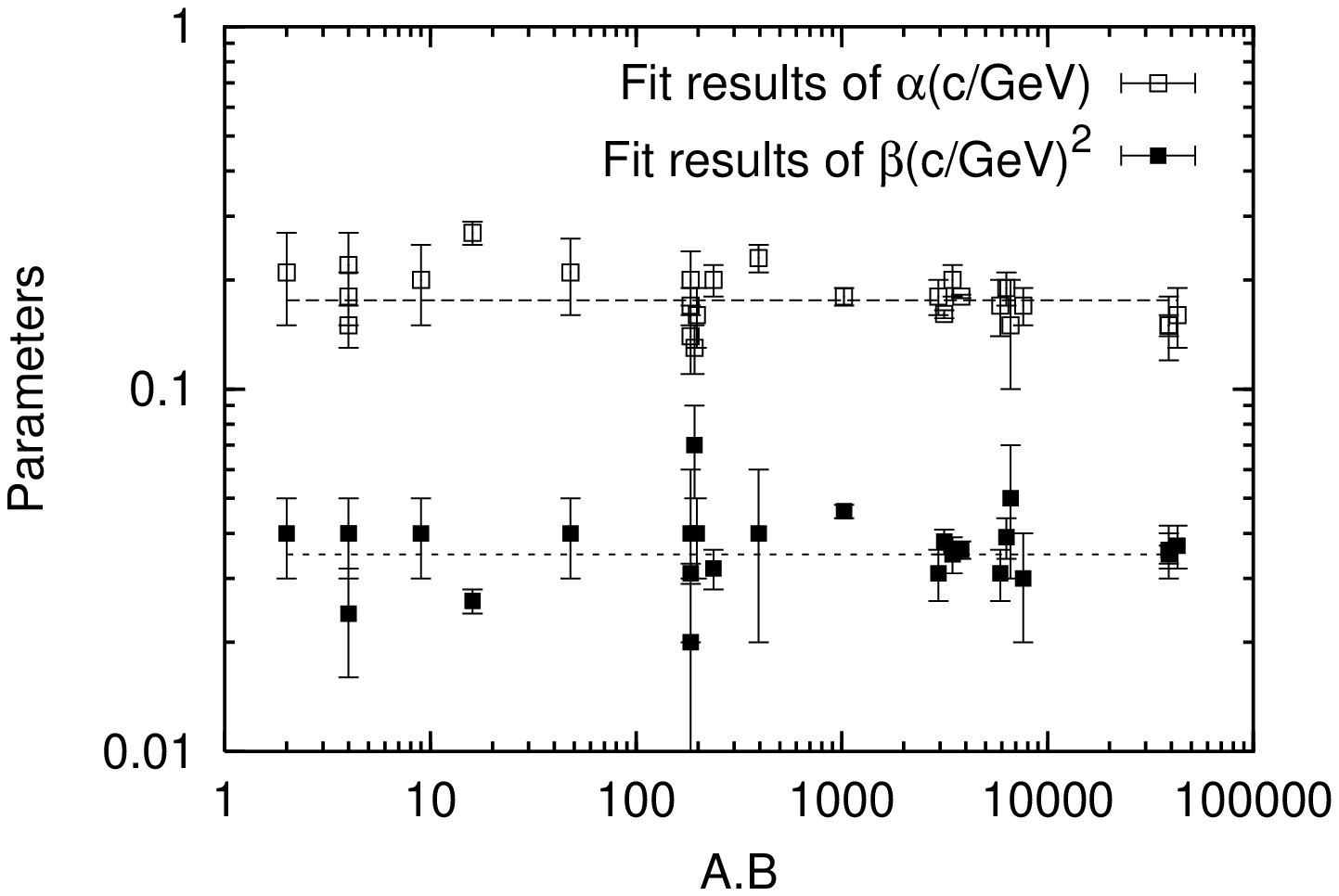}
\label{Fig.10} \caption{Values of $\alpha$ and $\beta$ for
different collisions as functions of the product of mass
numbers($AB$) of the interacting nuclei. The fitted values of
$\alpha$ and $\beta$, enlisted in Table3, are taken as the data
points; and are denoted by empty and filled squares respectively.
The dashed line gives the average values $0.18\pm 0.03$ for
$\alpha$ and $0.035\pm 0.009$ for $\beta$.}
\end{figure}
\begin{figure}
\centering
\includegraphics[width=8.5cm]{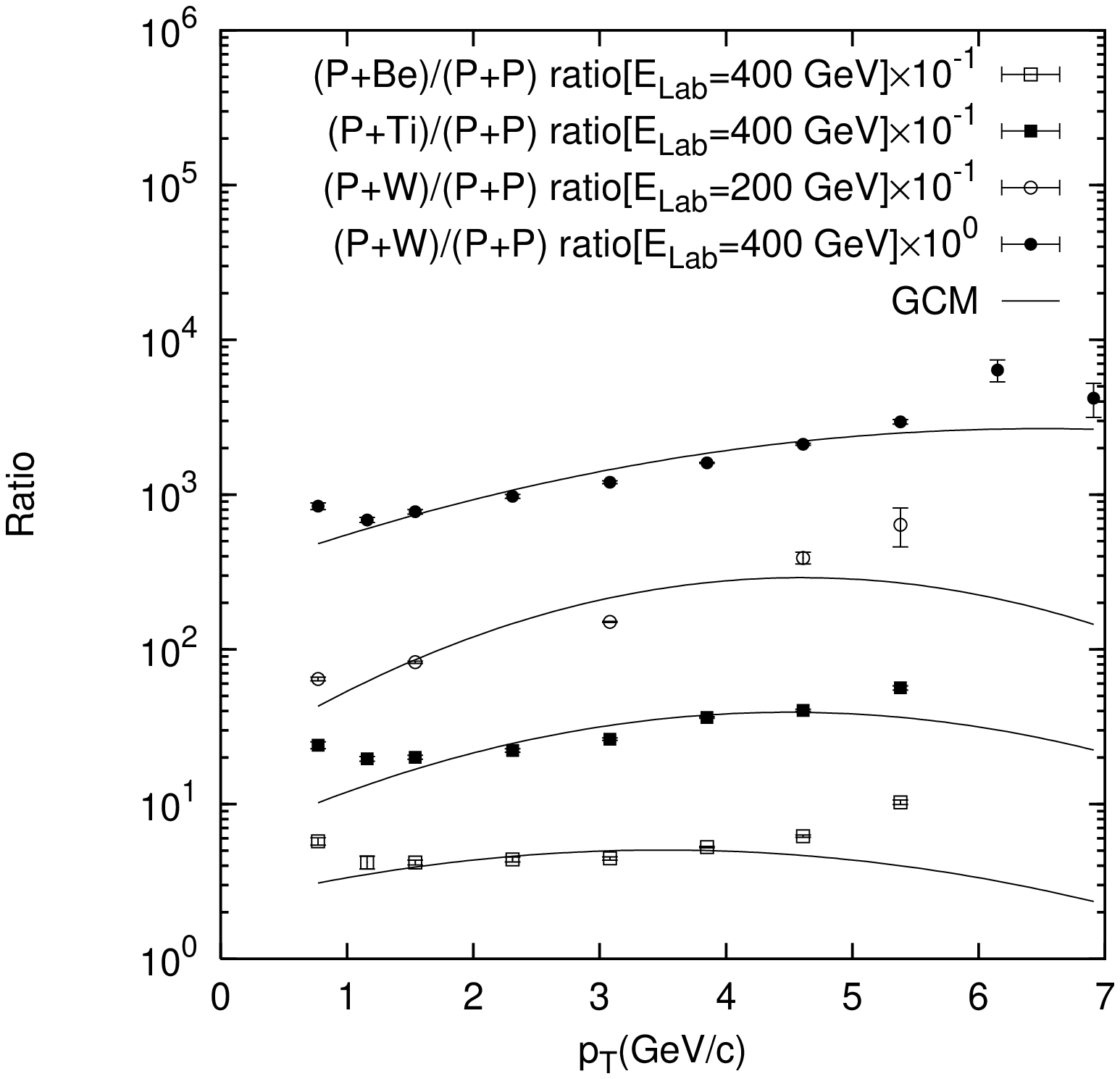}
\label{Fig.11} \caption{Plot of ratios of cross-sections for
different Proton-induced collisions at different energies to that
for Proton-Proton collisions at the corresponding energies. The
data-points are taken from Ref.\cite{Antreasyan1}. The present model-based results
are shown by the solid curves.}
\includegraphics[width=8.5cm]{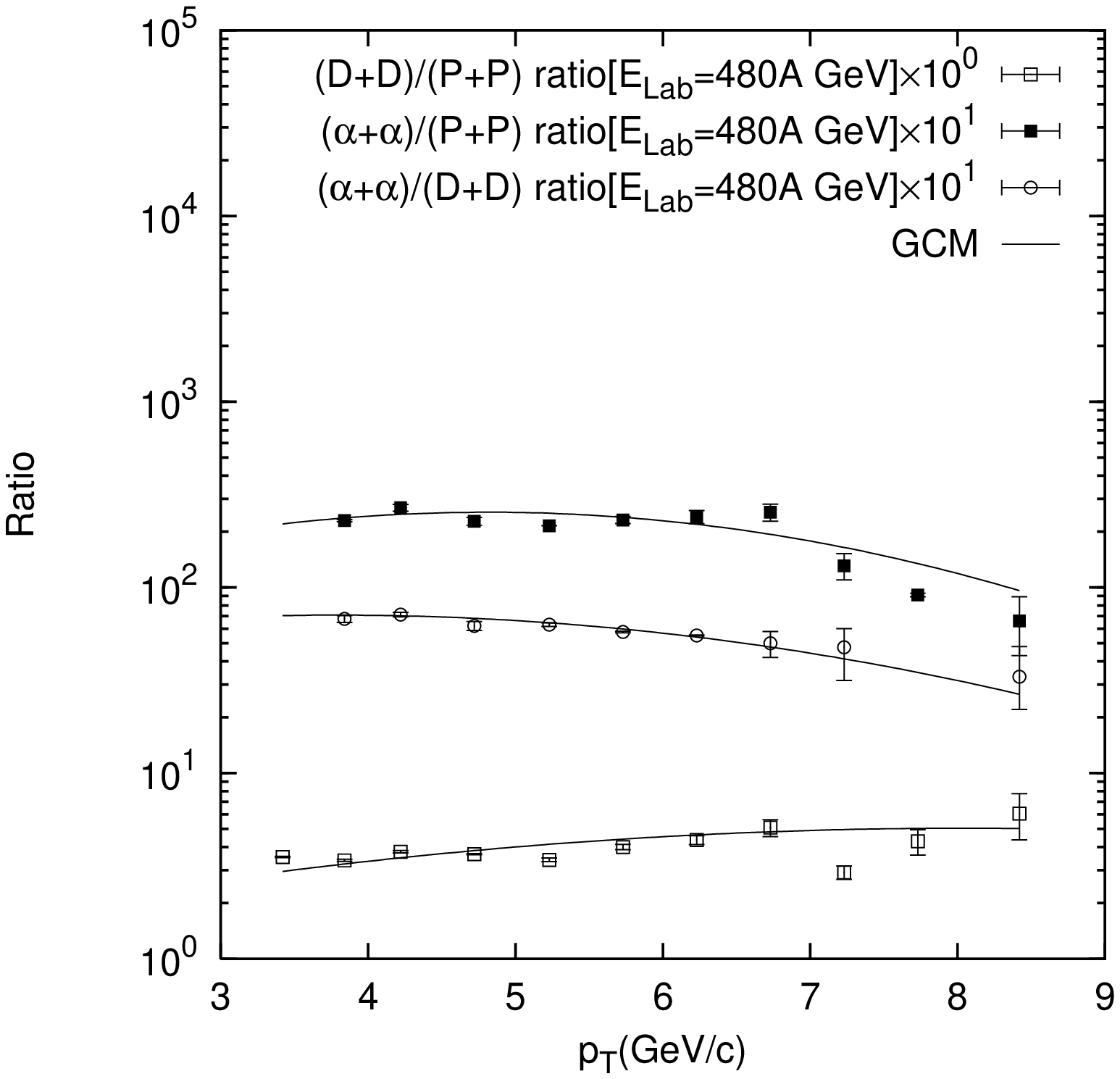}
\label{Fig.12} \caption{Plot of some cross section ratios as a
function of $p_T$. The data points are taken
from Ref.\cite{Angelis1}. The solid curves depict the present model-based
results.}
\end{figure}
\begin{figure}
\centering
\includegraphics[width=8.5cm]{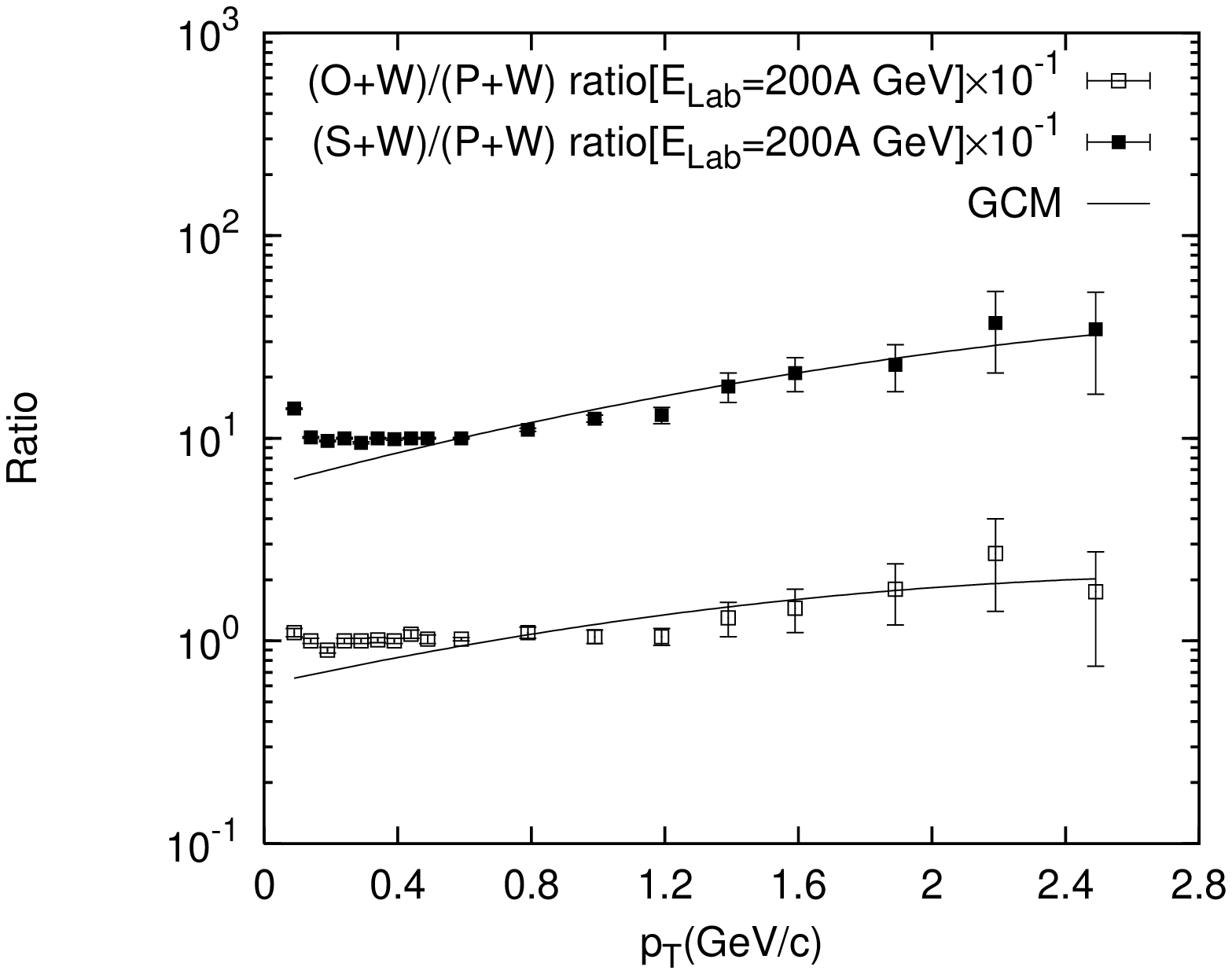}
\label{Fig.13} \caption{The nature of cross section ratios for
 $\frac{O+W}{P+W}$ and
$\frac{S+W}{P+W}$ with respect to transverse momentum $p_T$. The
data points are taken from Ref.\cite{Akesson2,Donaldson1}. The present model-based results are shown by the solid curvilinear lines.}
\includegraphics[width=8.5cm]{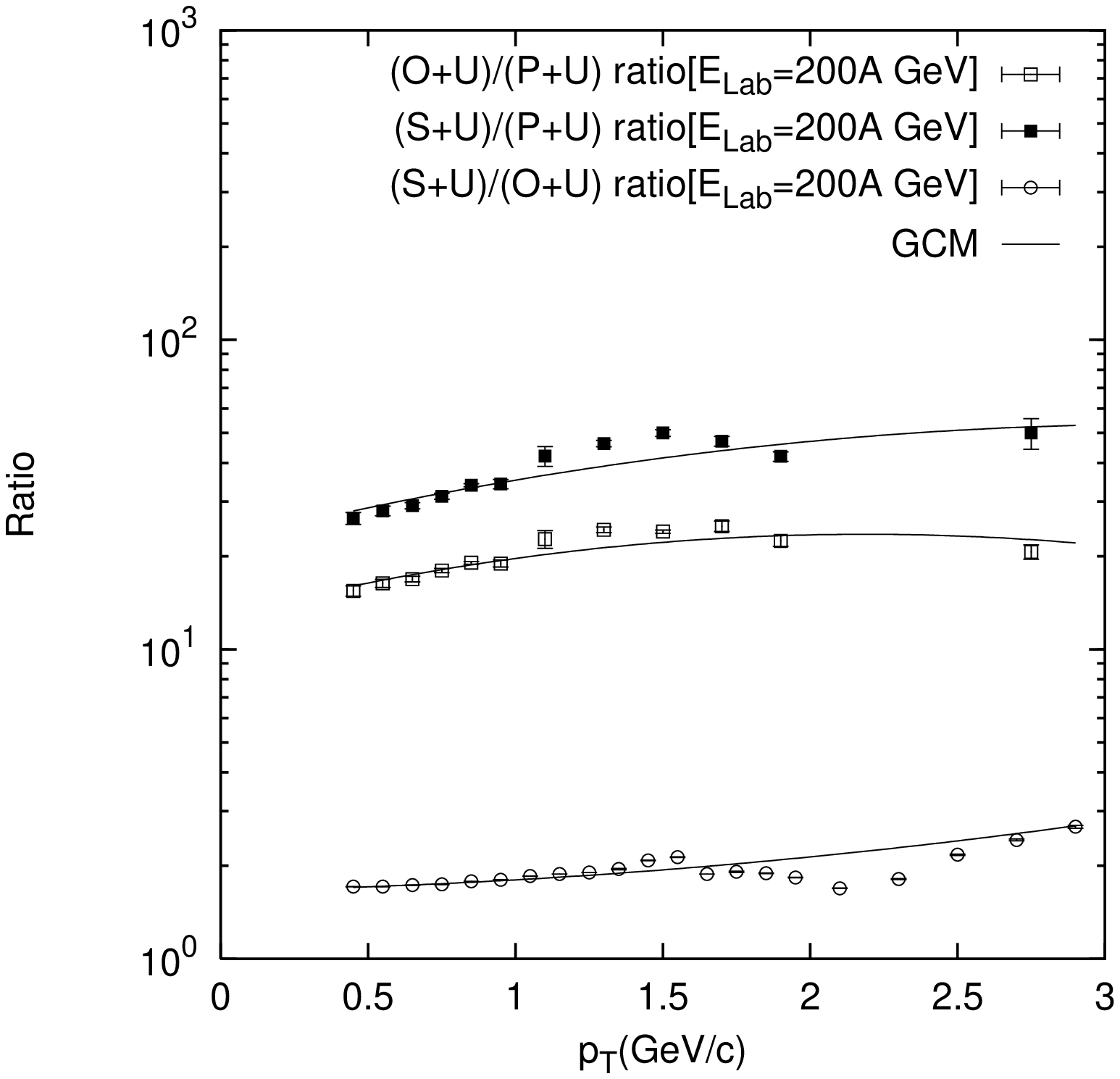}
\label{Fig.14} \caption{Plot of $\frac{O+U}{P+U}$,
$\frac{S+U}{P+U}$ and $\frac{S+U}{O+U}$ cross section ratios with
respect to transverse momentum $p_T$. The data-points are taken from Ref.\cite{Abreu1}. The solid curves provide the present
model-based results.}
\end{figure}
\begin{figure}
\centering
\includegraphics[width=8.5cm]{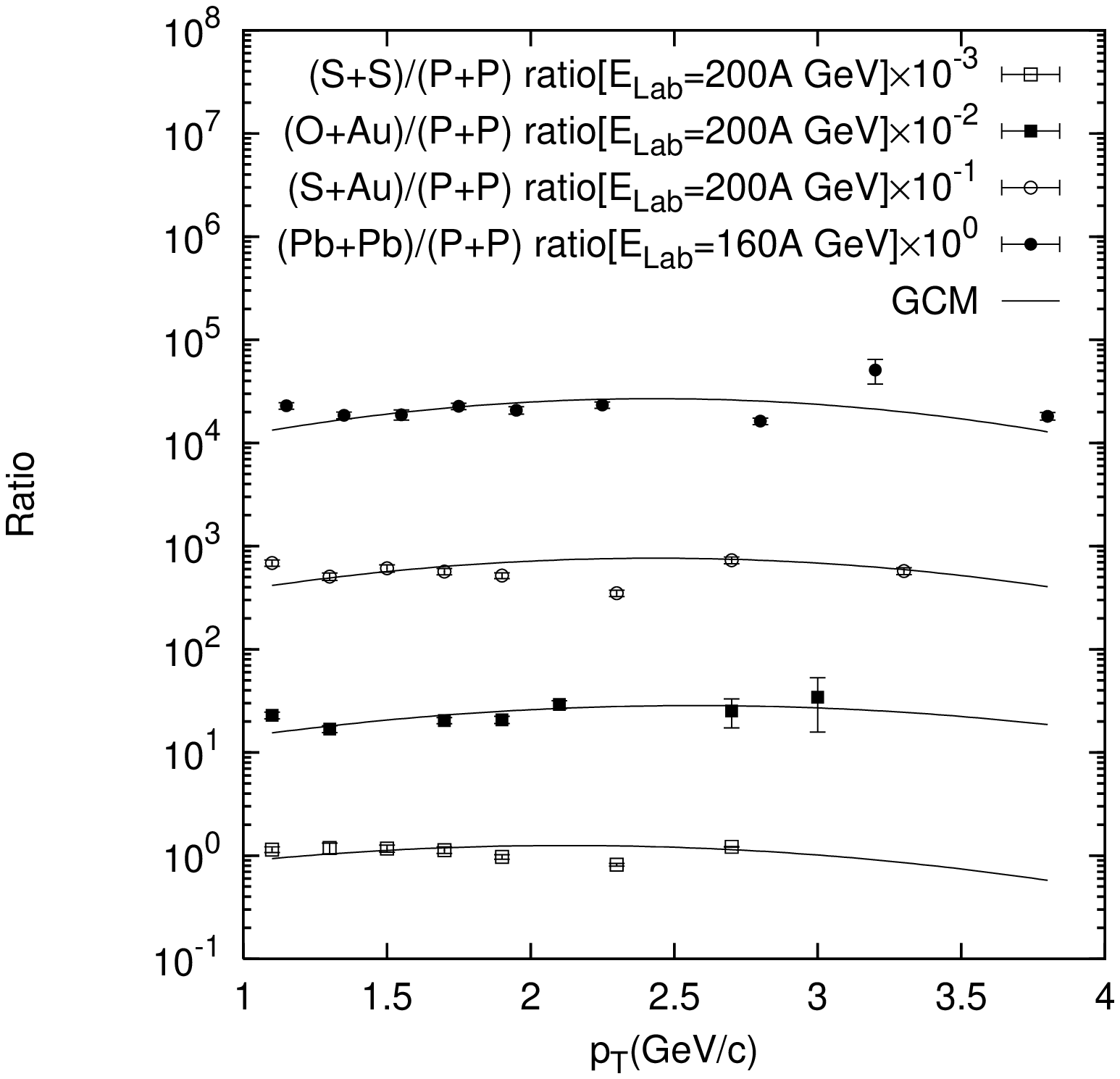}
\label{Fig.15} \caption{Plot of ratios of cross-sections for
different Nucleus-Nucleus collisions at different energies to that
for Proton-Proton collisions at the corresponding energies. The
data points are taken from Ref.\cite{Albrecht1,Aggarwal1,Albrecht2,Donaldson1}. To extract the $\frac{Pb+Pb}{P+P}$ ratio-data, we have divided the data for $Pb+Pb$ collision at $160A$ GeV by those for $P+P$ collision at $200A$ GeV. And this is an approximation. The model-based results of the present work are shown by the solid curves.}
\subfigure[]{
\begin{minipage}{.5\textwidth}
\includegraphics[width=8cm]{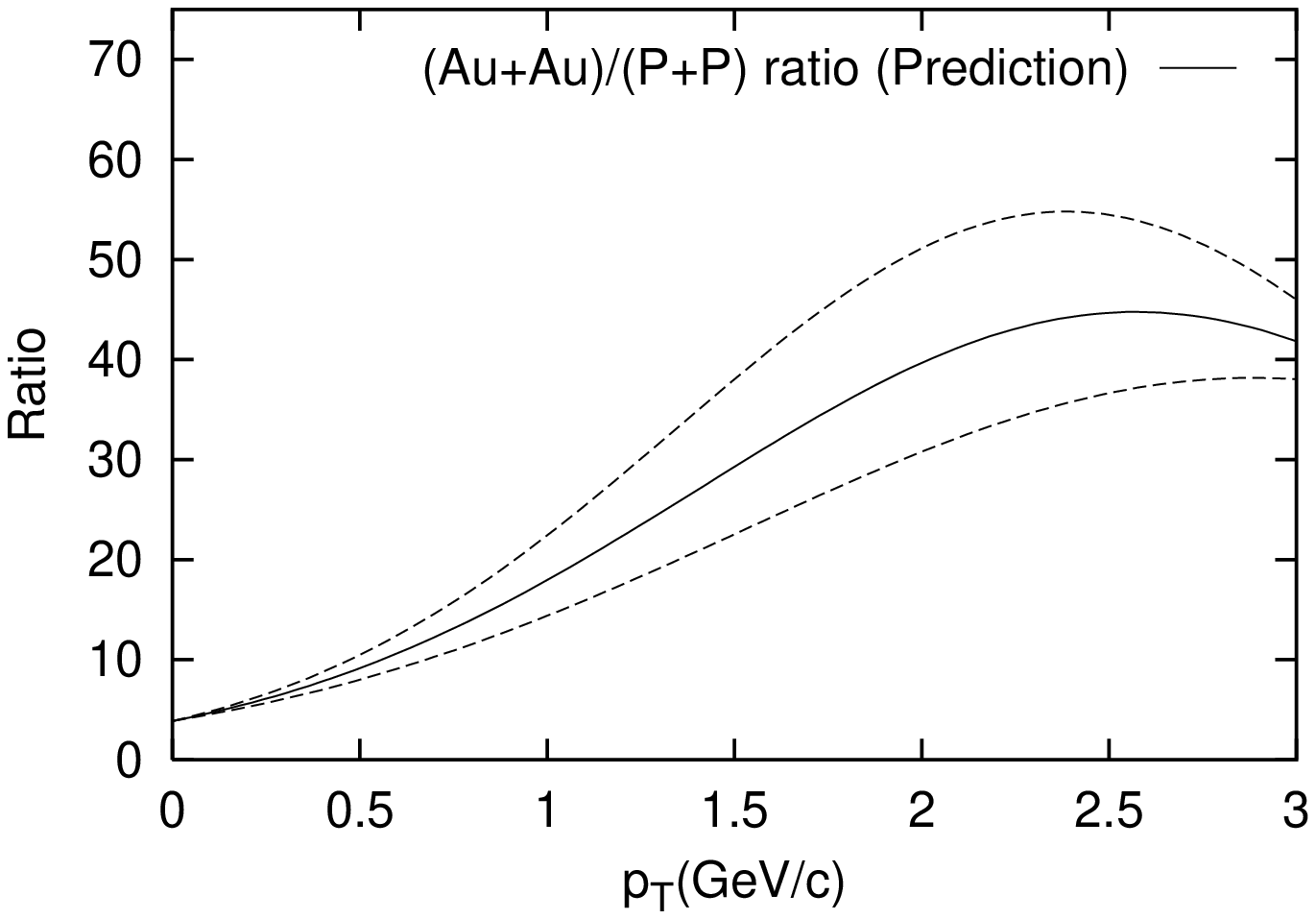}
\label{Fig.16a}
\end{minipage}}%
\subfigure[]{
\begin{minipage}{.5\textwidth}
\centering
\includegraphics[width=8cm]{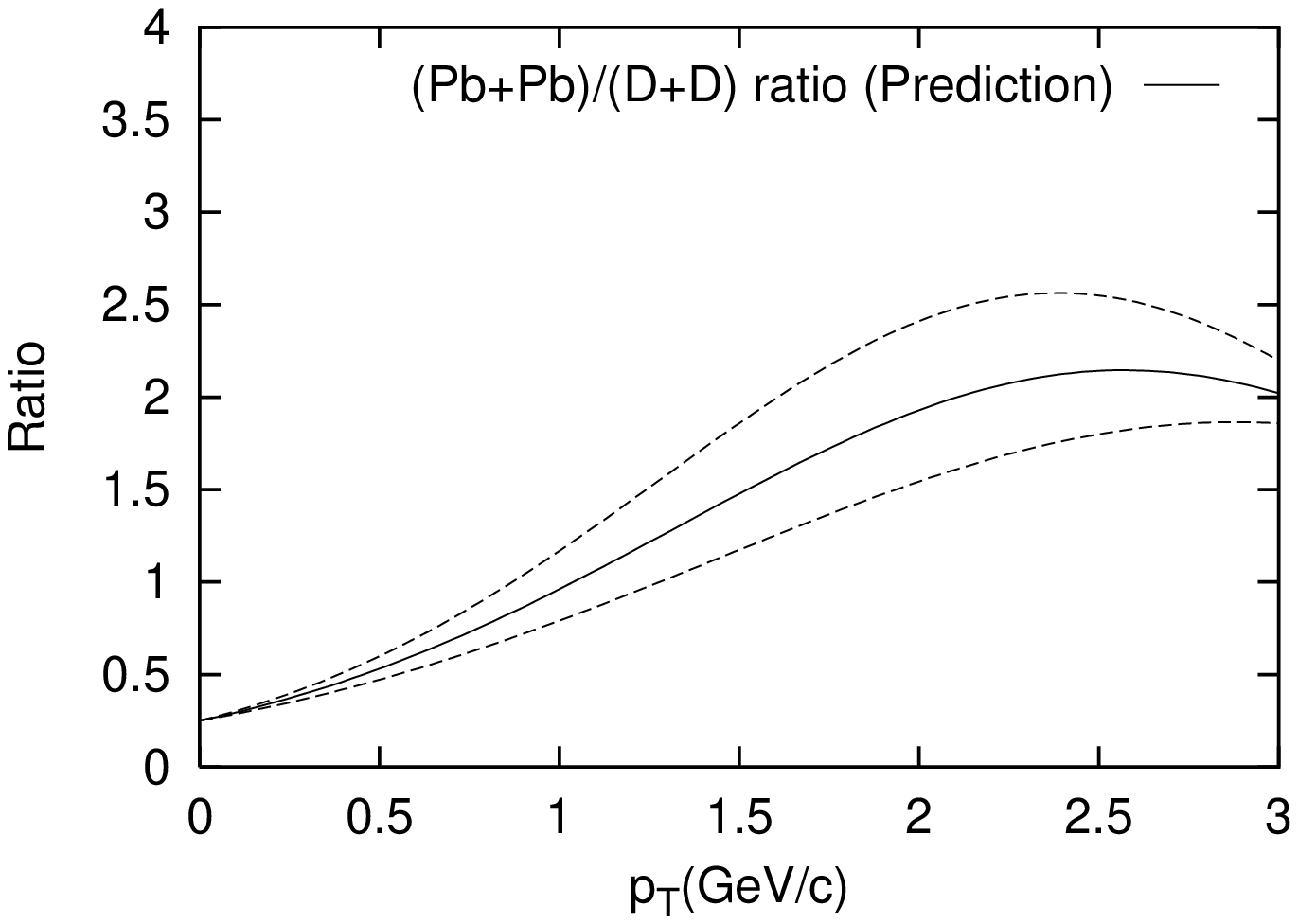}
\label{Fig.16b}
\end{minipage}}%
\caption{Predictive nature of $\frac{Pb+Pb}{D+D}$ and $\frac{Au+Au}{P+P}$ cross section ratio with respect to the transverse momentum($p_T$) on the basis of present model. The values of the ratio are certainly not exact, unless the $\epsilon$-value for pion production in corresponding reactions at any definite energy could be ascertained. The dotted curves depict the uncertainties arising out of the errors in $\alpha$ and $\beta$.}
\end{figure}
\begin{figure}
\subfigure[]{
\begin{minipage}{1\textwidth}
\centering
\includegraphics[width=8.5cm]{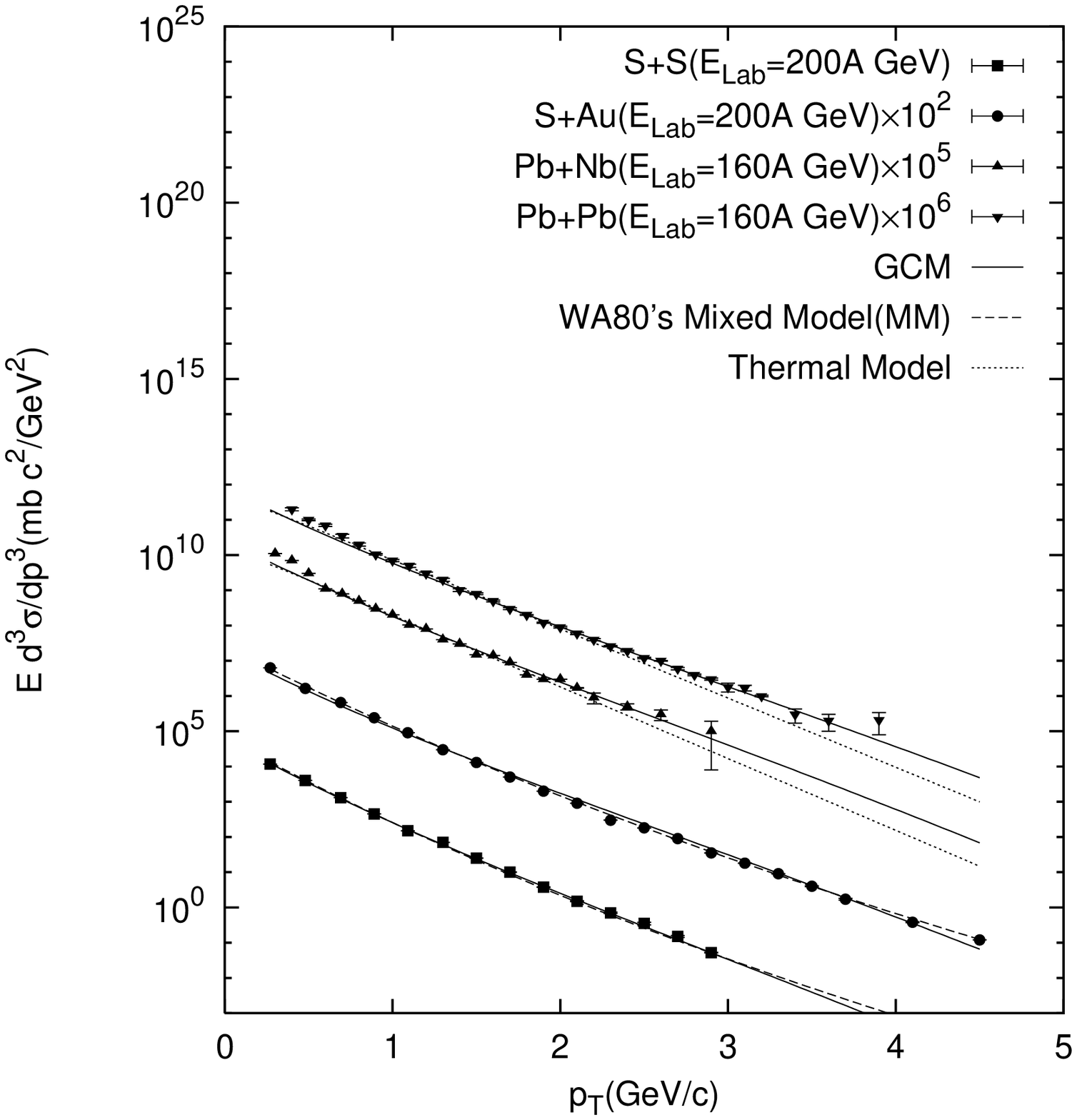}
\label{Fig.17a}
\end{minipage}}%
\vspace{.1cm}
\subfigure[]{
\begin{minipage}{1\textwidth}
\centering
\includegraphics[width=8.5cm]{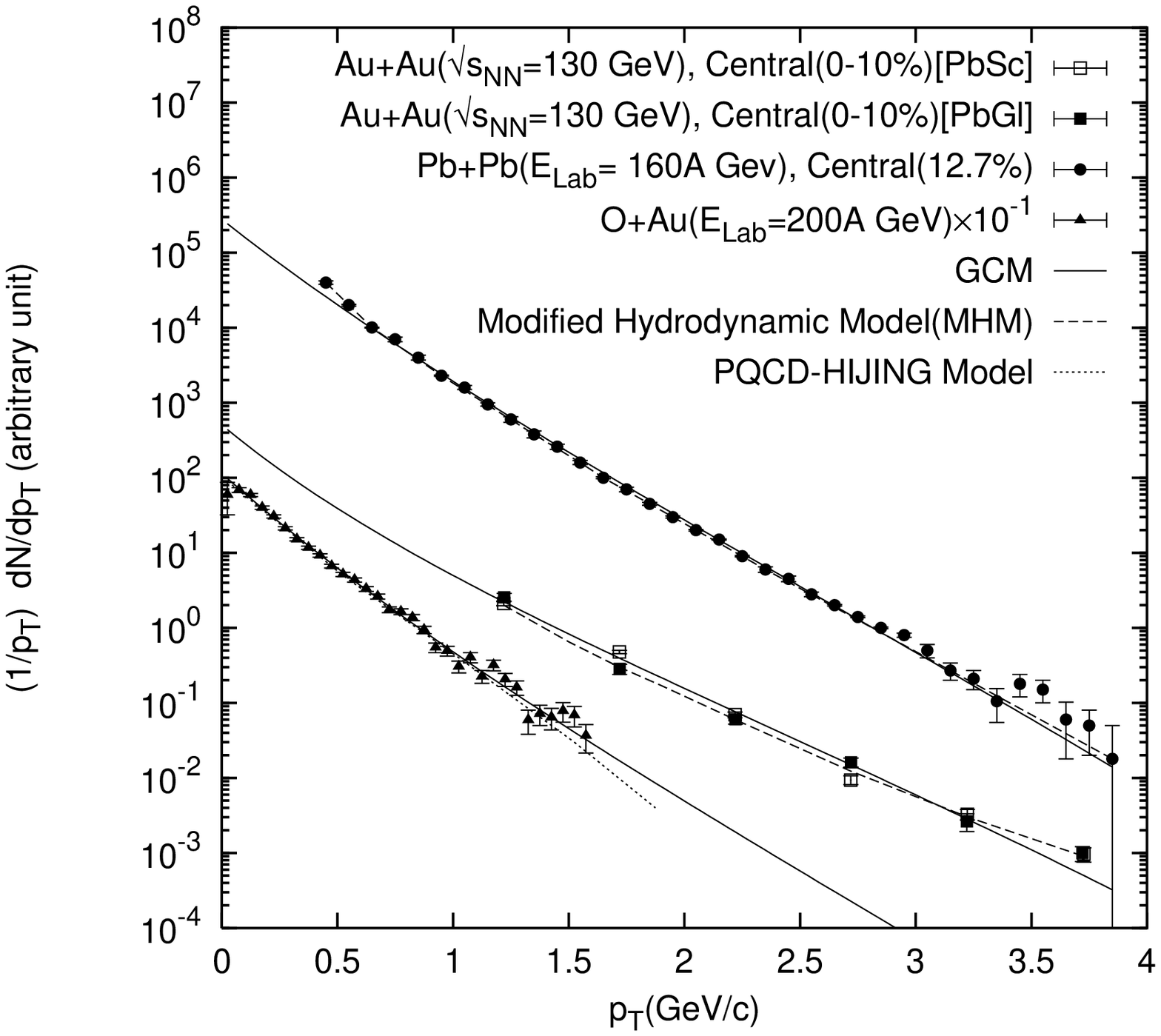}
\label{Fig.17b}
\end{minipage}}%
\caption{Comparison of results obtained on the basis of various theoretical models with those attained by GCM against measured data on pion production in some selected nucleus-nucleus collisions. For Fig.(17a) the data sources for $SS$ and $SAu$, for $PbPb$ and $PbNb$ are \cite{Albrecht1} and \cite{Aggarwal1} respectively. The theoretical plots based on Mixed Model and Thermal model are found from \cite{Albrecht1} and \cite{Aggarwal1}. In Fig.17(b) the data on $AuAu$ and $PbPb$ are obtained from \cite{Adcox1} and \cite{Aggarwal1}. The source of the reports on measurements for $OAu$ reaction is Ref.\cite{Alber1}. The theoretical plots based on MHM and on PQCD-HIJING model are assorted form
\cite{Peitzmann2} and \cite{Wang2}.}
\end{figure}

\section*{References}
\begin{thebibliography}{*}
\bibitem{McLerran1} McLerran L and Schuffner-Bielich J 2001 {\it{Phys. Lett.}} {\bf B
514} 29.
\bibitem{Pirner1} Pirner H J and Yuang F 2001 {\it{Phys. Lett.}} {\bf B
512} 297.
\bibitem{Adler1} Adler C \etal, STAR Collaboration 2001 {\it{Phys. Rev. Lett.}}
{\bf 87} 112303.
\bibitem{Wang1} Wang E and Wang X 2001 {\it{Phys. Rev.}} {\bf C 64} 034901.
\bibitem{Wong1} Wong C Y 1986 {\it{Phys. Rev.}} {\bf D 33} 711;
Hao S B and Wong C Y 1985 {\it{Phys. Rev.}} {\bf D 32} 1706.
\bibitem{Zhen1} Zhen B P and Ming S Y 1996 {\it{Phys. Lett.}} {\bf B 375} 355.
\bibitem{De1} De Bhaskar, Bhattacharyya S and Guptaroy P {\it{Int. Jour. Mod. Phys.}} {\bf A}(to be published).
\bibitem{Hagedorn1} Hagedorn R 1965 {\it{Nuovo. Cim. Suppl.}} {\bf 3} 147;
ibid 1968 {\bf 6} 169.
\bibitem{Hagedorn2} Hagedorn R 1983 {\it{Riv. Nuovo. Cim.}} {\bf 6} 1.
\bibitem{Faessler1} Faessler M A 1984 {\it{Phys. Rep.}} {\bf 115} 1.
\bibitem{Peitzmann1} Peitzmann T 1999 {\it{Phys. Lett.}} {\bf B 450} 7.
\bibitem{Schmidt1} Schmidt H R and Schukraft J 1993 {\it{J. Phys.}} {\bf G 19} 1705.
\bibitem{Bielich1} Bielich J S \etal 2002 {\it{Nucl. Phys.}} {\bf A 705} 494.
\bibitem{Albrecht1} Albrecht R \etal, WA80 Collaboration 1998 {\it{Eur. Phys. Jour.}} {\bf C 5} 255.
\bibitem{Antreasyan1} Antreasyan D \etal 1979 {\it{Phys. Rev.}} {\bf D 19} 764.
\bibitem{Aggarwal1} Aggarwal M M \etal, WA98 Collaboration 2002 {\it{Eur. Phys. Jour.}} {\bf C 23} 225.
\bibitem{David1} David G, PHENIX Collaboration 2002 {\it{Nucl. Phys.}} {\bf A 698} 227.
\bibitem{Adcox1} Adcox K \etal, PHENIX Collaboration 2002 {\it{Phys. Rev.
Lett.}} {\bf 88} 022301.
\bibitem{Akesson1} Akesson T \etal, NA34 Collaboration 1990 {\it{Z. Phys.}} {\bf C 46} 369; Drees A, NA34 Collaboration, 1989 {\it{Ph.D. thesis, Heidelberg University, Germany}}.
\bibitem{Peitzmann2} Peitzmann T 2002 {\it{nucl-th/0207012}}.
\bibitem{Wang2} Wang X N and Gyulassy M 1991 {\it{Phys. Rev.}} {\bf D 44}, 3501; Wang X N 1997 {\it{Physics Reports}} {\bf 280} 287.
\bibitem{Alper1} Alper B \etal 1975 {\it{Nucl. Phys.}} {\bf B 100} 237.
\bibitem{Drijard1} Drijard D \etal 1982 {\it{Nucl. Phys.}} {\bf B 208} 1.
\bibitem{Albajar1} Albajar C \etal 1990 {\it{Nucl. Phys.}} {\bf B 335} 261.
\bibitem{Bocquet1} Bocquet J \etal 1996 {\it{Phys. Lett.}} {\bf B 366} 434. 
\bibitem{Abe1} Abe F \etal 1988 {\it{Phys. Rev. Lett.}} {\bf 61} 1819.
\bibitem{Albrecht2} Albrecht R \etal, WA80 Collaboration 1988 {\it{Z. Phys}}. {\bf C 38} 97.
\bibitem{Abreu1} Abreu M C \etal, NA38 Collaboration 1993 {\it{Z. Phys.}} {\bf C 55} 365.
\bibitem{Angelis1} Angelis A L S \etal, BCMOR Collaboration 1987 {\it{Phys. Lett.}} {\bf B 185} 213.
\bibitem{Clark1} Clark A G \etal 1978 {\it{Nucl. Phys.}} {\bf B 142} 189.
\bibitem{Akesson2} Akesson T \etal, HELIOS Collaboration 1990 {\it{Z. Phys.}} {\bf C 46} 361.
\bibitem{Alber1} Alber T \etal 1998 {\it{Eur. Phys. Jour.}} {\bf C 2} 643.
\bibitem{Wlodarczyk1} Wlodarczyk Z  {\it{Proc. of the xxiii ICRC, Invited, Rapporteur
$\&$ Highlight Papers, (Calgary, Alberta, Canada, 19-30 July 1993)}}
Ed. D A Leahy, R B Hicks and D Venkatesan, {\it{Singapore: World Scientific}} p 355.
\bibitem{Agakichiev1} Agakichiev G \etal, CERES Collaboration 1998 {\it{Nucl. Phys.}} {\bf A 661} 23c.
\bibitem{Donaldson1} Donaldson G \etal 1976 {\it{Phys. Rev. Lett.}} {\bf 36} 1110.
\end{thebibliography}
\end{document}